\documentclass[acmtog]{acmart}


\usepackage{array}
\usepackage{color}
\usepackage{color}
\definecolor{green}{rgb}{0, 0.5, 0}
\definecolor{orange}{rgb}{0.8, 0.6, 0.2}
\definecolor{red}{rgb}{1.0, 0.0, 0.0}
\definecolor{teal}{rgb}{0.0, 0.4, 0.4}
\definecolor{purple}{rgb}{0.65,0,0.65}
\definecolor{saffron}{rgb}{0.95,0.75,0.2}
\definecolor{turquoise}{rgb}{0.0,0.5,0.5}

\usepackage[ruled,vlined,linesnumbered]{algorithm2e}
\usepackage{multirow}

\newcommand{\kx}[1]{{\color{black}#1}}
\newcommand{\new}[1]{{\color{black}#1}}
\newcommand{\wxg}[1]{{\color{black}#1}}
\newcommand{\wang}[1]{{\color{black}#1}}

\usepackage[normalem]{ulem}

\usepackage{mathtools}
\usepackage{amsmath, amsthm, amssymb, amsfonts, amsopn}
\usepackage{graphicx} % Necessary to use \scalebox
\usepackage{textcomp}
\usepackage{xfrac}
\usepackage{bbm}

\usepackage{overpic}
\usepackage{subfig}
\usepackage{wrapfig}

\newcommand{\hidecomment}[1]{}
\setcopyright{none}

\setcitestyle{square}

\begin{document}

\setcopyright{acmcopyright}
\acmJournal{TOG}
\acmYear{2018}\acmVolume{37}\acmNumber{6}\acmArticle{1}\acmMonth{11}
\acmDOI{10.1145/3272127.3275009}

% Title portion
%\title{Semantic Labelling of Raw 3D CAD Shape via Part Hypotheses}
\title{Learning to Group and Label Fine-Grained Shape Components}

\author{Xiaogang Wang}
\affiliation{%
	\institution{Beihang University}}
\author{Bin Zhou}
\affiliation{%
	\institution{Beihang University}}
\author{Haiyue Fang}
\affiliation{%
	\institution{Beihang University}}
\author{Xiaowu Chen}
\affiliation{%
	\institution{Beihang University}}
\author{Qinping Zhao}
\affiliation{%
	\institution{Beihang University}}
\author{Kai Xu}
\authornote{Corresponding author: Kai Xu (kevin.kai.xu@gmail.com)}
\affiliation{%
	\institution{National University of Defense Technology and Princeton University}}

\begin{abstract}
\new{A majority of stock 3D models in modern shape repositories are assembled
with many fine-grained components.
The main cause of such data form is the component-wise modeling process widely practiced by human modelers. These modeling components thus inherently reflect some function-based shape decomposition
the artist had in mind during modeling. On the other hand, modeling components represent
an over-segmentation since a functional part is usually modeled as a multi-component assembly.
Based on these observations, we advocate that labeled segmentation of stock 3D models
should not overlook the modeling components and propose a learning solution
to grouping and labeling of the fine-grained components.
However, directly characterizing the shape of individual components for the purpose of
labeling is unreliable, since they can be arbitrarily tiny and semantically meaningless.
\wang{
We propose to generate part hypotheses from the components based on a hierarchical grouping strategy,
and perform labeling on those part groups instead of directly on the components.
}
Part hypotheses are mid-level elements which are more probable to carry semantic information.
A multi-scale 3D convolutional neural network is trained to extract context-aware features for the hypotheses.
To accomplish a labeled segmentation of the whole shape, we
formulate higher-order conditional random fields (CRFs) to infer
an optimal label assignment for all components.
Extensive experiments demonstrate that our method achieves significantly robust labeling results
on raw 3D models from public shape repositories.
Our work also contributes the first benchmark for component-wise labeling.}
\end{abstract}


\begin{teaserfigure}
  \includegraphics[width=\textwidth]{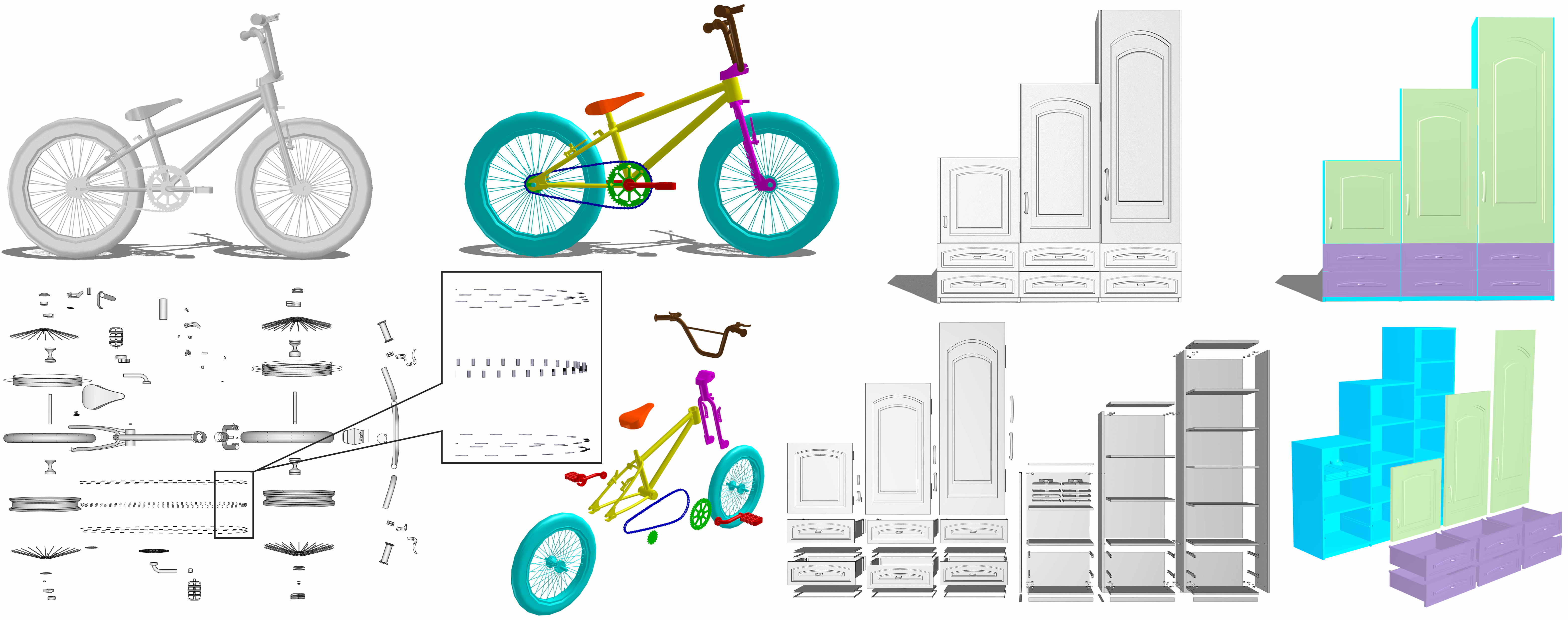}
  \caption{\new{We study a novel problem of semantic labeling of raw 3D models from online shape repositories, composed of highly fine-grained components.
Given a 3D shape comprising numerous human-crafted components,
our method produces high quality labeling result (indicated by distinct colors). This is achieved by aggregating the components into \emph{part hypotheses} and characterizing these mid-level elements for robust grouping and labeling of fine-grained components.}}
  \label{fig:teaser}
\end{teaserfigure}

\maketitle

\section{Introduction}
%Motivation
\new{Semantic or labeled segmentation of 3D shapes has gained significant performance boost
over recent years,
benefiting from the advances of machine learning techniques~\cite{Kalogerakis_SG10,Hu_SGP12},
and more recently of deep neural networks~\cite{Kalogerakis_CVPR17,Su_CVPR17}.
Existing methods have so far been dealing with manifold meshes, point clouds,
or volumes. They are, however, not specifically designed to handle most stock 3D models,
which typically assembles
up to hundreds of highly fine-grained components (Figure~\ref{fig:teaser}).
Multi-component assembly is the most commonly seen data form
in modern 3D shape repositories (e.g., Trimble 3D Warehouse~\cite{Tri3Dwarehouse} and ShapeNet~\cite{ShapeNet2015}).
See Figure~\ref{fig:shapenetstat}(left) for the statistics of component counts in ShapeNetCore.

Multi-view projective segmentation~\cite{Wang_SA13,Kalogerakis_CVPR17} is perhaps the most feasible approach
for handling multi-component shapes, among all existing techniques.
View-based methods are representation independent, making them applicable to non-manifold
models. However, a major drawback of this approach is
that it cannot handle shapes with severe self-occlusion.
Components hidden from the surface are invisible to any view,
thus cannot be labeled.
Figure~\ref{fig:car}(a) shows such an example: The seats in the car are completely occluded by
the car shell and thus cannot be segmented or labeled correctly by view-based methods.

Most off-the-shelf 3D models are created by human modelers in a component-by-component fashion.
Generally, human modelers tend to have in mind a meaningful decomposition of the target object before starting.
Such decomposition is inherently related to functionality, mimicking the actual production of the
man-made objects, e.g., a car is decomposed into shell, hood, wheels, seats, etc.
Therefore, we advocate that the segmentation of such models should not overlook the components
coming with the models.
Meanwhile, these components usually represent an over-segmentation -- a functional part
might be modeled as an assembly of multiple sub-parts.
A natural solution to semantic segmentation thus seems to be a labeled grouping of the modeling components.

\begin{figure}[t]
  \centering
  \includegraphics[width=\linewidth]{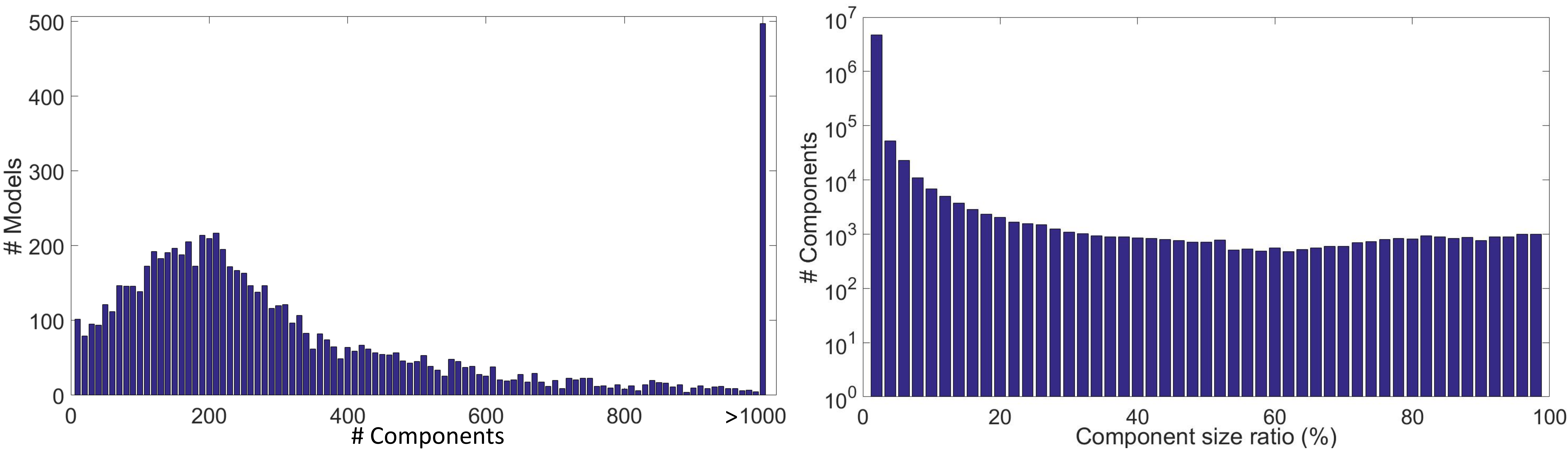}
  \caption{\new{Statistics on component count (left) and size (right) of the models in ShapeNetCore. The left histogram is measured only for car models. In the right one, size is measured by the ratio of bounding box volume between a component and the whole shape.}}
  \label{fig:shapenetstat}
\end{figure}

A few facts about the components of stock models, however, make their grouping and labeling especially difficult.
\emph{First}, the decomposition of these models is often highly fine-grained.
See the tiny components the bicycle model in Figure~\ref{fig:teaser} contains.
%In Figure~\ref{fig:ShapeNetCore_stati_car}, we provide a statistics on the number of components of
%all car models in ShapeNetCore.
Taking the car models in ShapeNetCore for example, about $85\%$ contains over $100$ components.
\emph{Second}, the size of components varies significantly; see Figure~\ref{fig:shapenetstat}(right).
\emph{Third}, different modelers may have different opinions about shape composition,
making the components of the same functional part highly inconsistent across different shapes.
The example in Figure~\ref{fig:car}(b) shows that the wheel parts from different vehicle models
have very different composition.
Due to these reasons, it is very unreliable to directly characterize the shape of individual components for the purpose of labeling.

These facts motivate us to consider larger and more meaningful elements, for achieving a robust semantic
labeling of fine-grained components.
In particular, we propose to generate \emph{part hypotheses} from the components, representing potential
functional or semantic parts of the object. This is achieved by a series of effective grouping
strategies, which is proven robust with extensive evaluation.
\wang{Our task then becomes labeling the true part hypotheses while pruning those false ones,
instead of directly labeling the individual components.}
Working with part hypotheses enables us to learn more informative shape representation,
based on which reliable labeling can be conducted.
Part hypothesis is similar in spirit to mid-level patch for image understanding which admits
more discriminative descriptors than feature points~\cite{Singh2012}.
%Different from mesh segmentation, however,
%our part hypotheses are generated by grouping human-crafted, more meaningful components,
%rather than triangles from scratch, enjoying the existing decompositions of stock models.

\begin{figure}[t]
  \centering
  \includegraphics[width=0.9\linewidth]{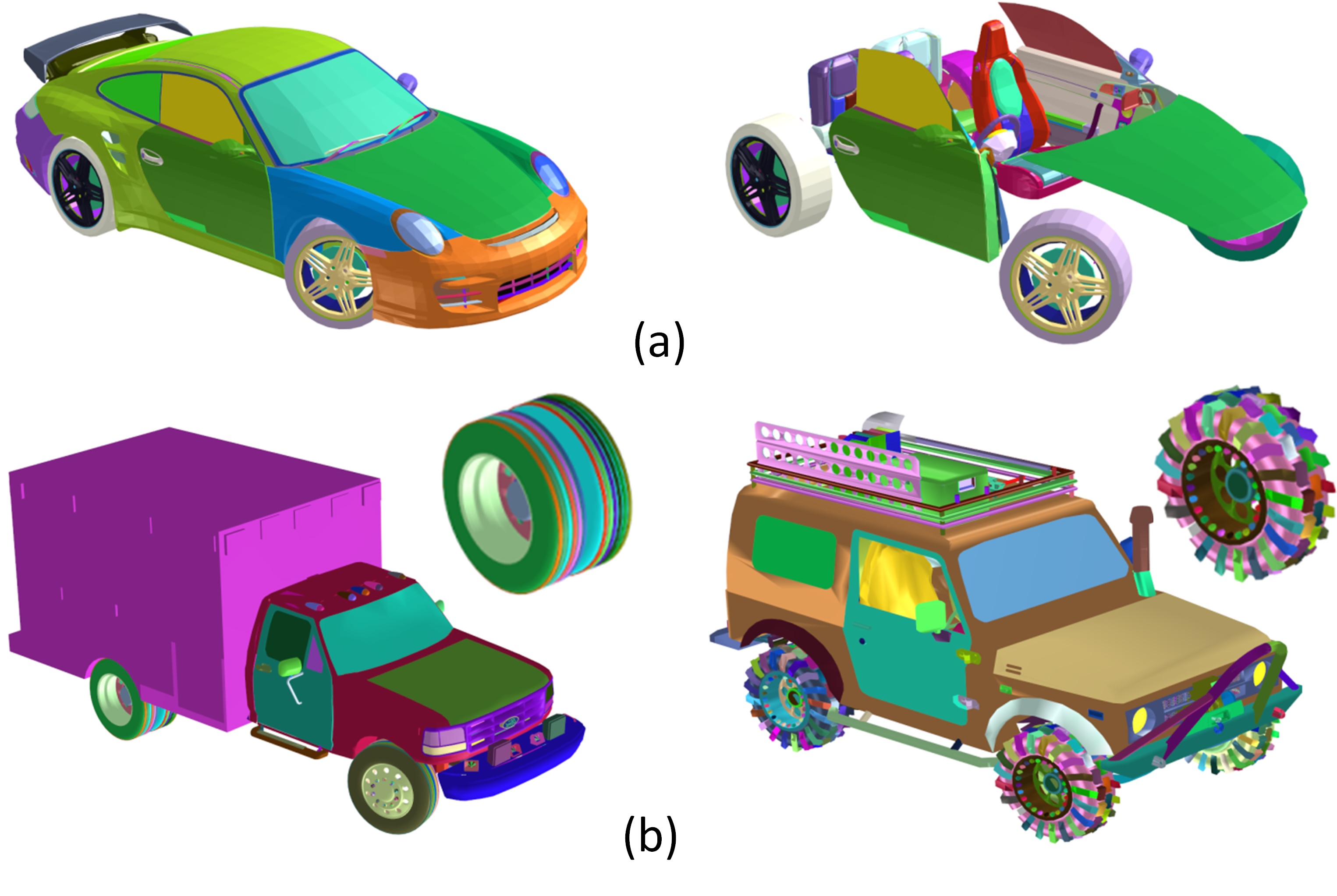}
  \caption{\new{Hidden components, e.g., car seats in (a), and various fine-grained decompositions of vehicle wheels (b)
  found in the ShapeNet.}}
  \label{fig:car}
\end{figure}

% Our method new insights 2:  multi-scale feature representation for part
To achieve a powerful part hypotheses labeling, we adopt 3D Convolutional Neural Networks (CNN)
to extract features from the volumetric representation of part hypotheses.
In order to learn features that capture not only local part geometry but also global, contextual information,
we design a network that takes two scales of 3D volume as input.
The local scale encodes the part hypothesis of interest itself, through feature extraction over
the voxelization of the part within its bounding box.
The global volume takes the bounding box of the whole shape as input, and encodes the context
with two channels contrasting the volume occupancy of the part hypothesis itself and that of the remaining parts.
The network outputs the labeling probabilities of the part hypothesis over different part categories, which are used for final labeled segmentation.

To accomplish a labeled segmentation of the whole shape,
we formulate higher-order Conditional Random Fields (CRFs) to
infer an optimal label assignment for each component.
%The data term is constructed based on the labeling probabilities of part hypotheses,
%while the smooth term regularizes the segmentation to produce as few as possible segments.
Our CRF-based model achieves highly accurate labeling, while saving the effort on preparing large amount of high-order relational data for training a deep model.
Consequently, our design choice, combining CNN-based part hypothesis feature and higher-order CRFs, achieves a good balance between model generality and complexity.

We validate our approach on our multi-component labeling (MCL) benchmark dataset.
The multi-component 3D models are collected from both ShapeNet and 3D Warehouse, with all
components manually labeled.
Our method achieves significantly higher accuracy in grouping and labeling
highly fine-grained components than alternative approaches. We also demonstrate how our method can be applied
to fine-grained part correspondence for 3D shapes, achieving state-of-the-art results.

%Our contributions
The main contributions of our paper include:
\begin{itemize}
  \item We study a new problem of labeled segmentation of stock 3D models based on the pre-existing, highly fine-grained components, and
  approach the problem with a novel solution of part hypothesis generation and characterization.
  \item We propose a multi-scale 3D CNN for encoding both local and contextual information for part hypothesis labeling, as well as a CRF-based formulation for component labeling.
  \item We build the first benchmark for multi-component labeling with component-wise ground-truth labels and conduct extensive evaluation over the benchmark.
\end{itemize}

\begin{figure*}[t]
  \centering
  \includegraphics[width=\textwidth]{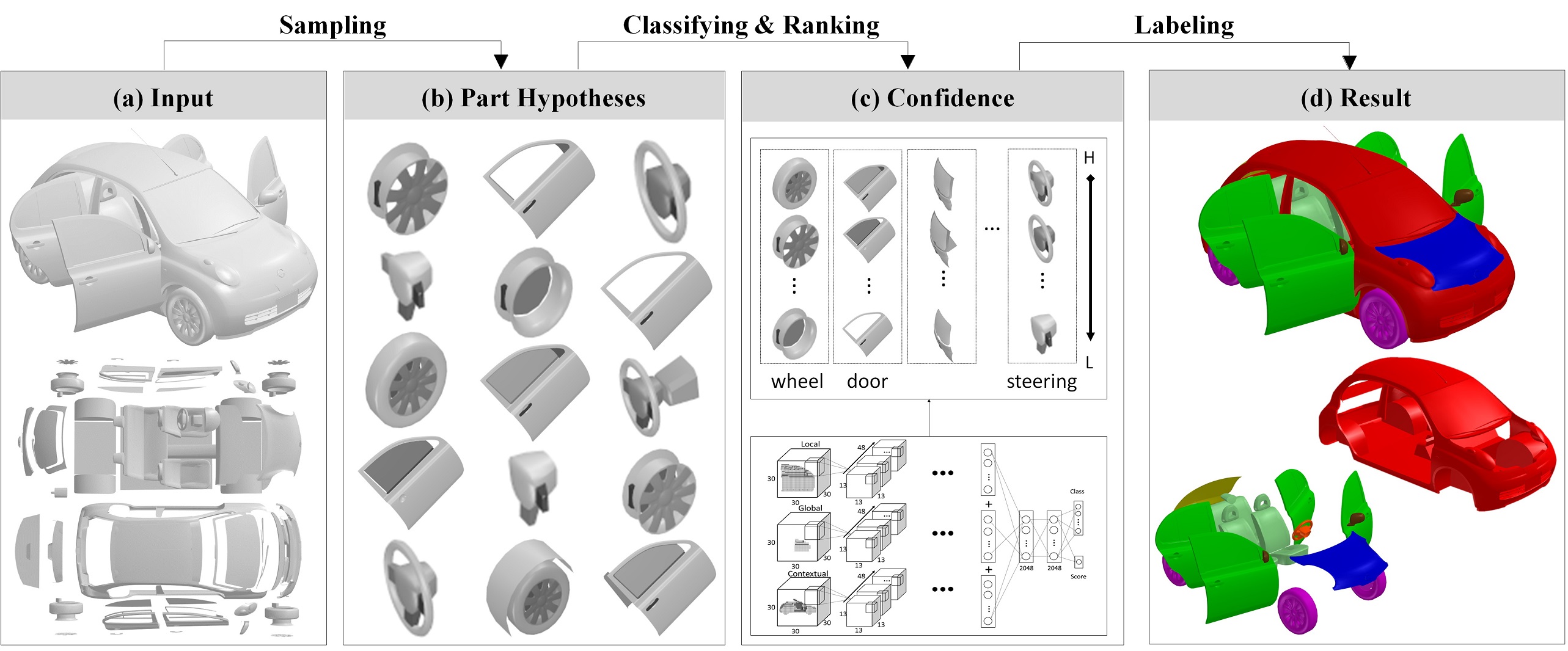}
  \caption{An overview of our approach.
Given a 3D model comprising many components (a),
our method first performs hierarchical sampling of candidate part hypotheses
with a bottom-up grouping procedure (b, Section~\ref{sec:sampling}).
Each candidate is then fed into a multi-scale Convolutional Neural Networks which predicts its label
and regresses a confidence score against each part label (c, Section~\ref{sec:ranking}).
Finally, the optimal label of each component is inferred with
a higher-order Conditional Random Fields, based on the confidence scores (d, Section~\ref{sec:labelling}).}
  \label{fig:pipeline}
\end{figure*}
}

\section{Related Work}
Shape segmentation and labeling is one of the most classical and long-standing problems
in shape analysis, with numerous methods having been proposed.
Early studies~\cite{Katz_SG03,Huang_EG09,Shapira_IJCV10,Zhang_TOG12,Au_TVCG12} most utilize
hand-crafted geometry features\wang{.}
%such as
%curvature~\cite{Gal_TOG06},
%PCA~\cite{Kalogerakis_SG10},
%shape diameter function~\cite{Shapira_IJCV10},
%distance from medial surface~\cite{Liu_EG09},
%average geodesic distance~\cite{Hilaga_SG01},
%shape context~\cite{Belongie_PAMI02},
%spin image~\cite{Johnson_PAMI99}, etc.
One geometric feature usually captures very limited aspects about shape decomposition and
a wider practiced approach is to combine multiple features~\cite{Kalogerakis_SG10}.

%Shape Segmentation and Labelling: Learning-based
To tackle the limitation of hand-crafted features,
data-driven feature learning methods are proposed~\cite{xu2016data}.
%and formulate the shape segmentation and labelling as a supervised classification problem to identify which part category each triangle belong to.
%Kalogerakis et al.~\shortcite{Kalogerakis_SG10} presented a method to segment and label 3D mesh
%by combining various geometric features of triangle(s)
%with the CRF framework and JointBoost classifier.
Guo et al.~\shortcite{Guo_TOG15} learned a compact representation
of triangle for 3D mesh labeling
by non-linearly combining and hierarchically compressing various geometry features with the deep CNNs.
Xie et al.~\shortcite{Xie_SGP14} proposed a fast method for 3D mesh segmentation and labeling
based on Extreme Learning Machine.
%They employed a set of feature descriptors and the Extreme Learning Machine (ELM) to
%train a neural networks classifier for mesh labelling.
Yi et al.~\shortcite{Yi_CVPR17} proposed a method, named SyncSpecCNN,
to label the semantic part of 3D mesh.
SyncSpecCNN trains vertex functions using CNNS,
and conducts spectral analysis to enable kernel weight sharing
by using localized information of mesh graphs.
These methods achieve promising performance,
while largely focusing on manifold and/or watertight surface mesh, but not suited for raw 3D models
from modern shape repositories.

%Shape Segmentation and Labelling: Projection/Image-based and PointCould
Recently,
%Wang et al.~\shortcite{Wang_SA13}
%treated a 3D shape as a collection of 2D projections,
%labelled each projection
%by transferring knowledge from existing labelled images,
%and back-projected and fused the part labellings on the 3D shape.
Kalogerakis et al.~\shortcite{Kalogerakis_CVPR17} proposed a deep architecture for
segmenting and labeling semantic parts of 3D shape
by combining multi-view fully convolutional networks and surface-based CRFs.
Projection-based methods~\cite{Wang_SA13,Kalogerakis_CVPR17} are suitable for imperfect (e.g., incomplete, self-intersecting, and noisy) 3D shapes, but inherently have a hard time on shapes with severe self-occlusion.
Su et al.~\shortcite{Su_CVPR17} designed a novel type of neural network, named PointNet,
for directly segmenting and labeling 3D point clouds while
respecting the permutation invariance, obtaining state-of-the-art performance on point data.

%Shape Co-analysis:
Several unsupervised or semi-supervised methods are proposed
for the co-segmentation and/or co-labeling of a collection of 3D shapes belonging to the
same category~\cite{xu2010style,Huang_SA11,Sidi_SA11,Wang_SA12,Hu_SGP12,Lv_PG12,van_SG13}.
Most of these methods are based on an over-segmentations of the input shapes.
A grouping process is then conducted to form semantic segmentation and labeling.
Such initial over-segments (e.g., superfaces) are analogy to our `part hypotheses'.
However, they are still too low level to capture meaningful part information.
Our method benefits from the pre-existing fine-grained components, which makes part
hypothesis based analysis possible.
%Huang et al.~\shortcite{Huang_SA11} segmented 3D meshes
%jointly utilizing the initial segments of individual shapes
%and possible correspondences between segments from multiple shapes.
%Sidi et al.~\shortcite{Sidi_SA11} clustered the initial segments
%in an embedded descriptor spectral space.
%Hu et al.~\shortcite{Hu_SGP12} performed co-segmenting
%using subspace clustering for the over-segmented patches.

%CAD Model Analysis
Few works studied on semantic segmentation of multi-component models.
Liu et al.~\shortcite{Liu_SA14} proposed to label and organize 3D scenes obtained from the Trimble 3D Warehouse
into consistent hierarchies capturing semantic and functional substructures.
The labeling is based on over-segmentation of the 3D input,
and guided by a learned probabilistic grammar.
Yi et al.~\shortcite{Yi_SG17} proposed a method of
converting the scene-graph of a multi-component shape into segmented parts
by learning a category-specific canonical part hierarchy.
Their method achieves fine-grained component labeling,
while scene graphs are not always available.

\section{Method}
\label{sec:method}
Please refer to Figure~\ref{fig:pipeline} for an overview of our algorithm pipeline.
In the next, we describe the three algorithmic components, including part hypothesis generation,
part hypothesis classification and scoring, and part composition inference and component labeling.

\subsection{Generating part hypothesis}
\label{sec:sampling}
\paragraph{\textbf{Part hypothesis}}
In our work, a semantic part, or part for short, refers to a semantically independent or functionally complete group of components. A part hypothesis is a component group which potentially represents a semantic part.
When searching for a part hypothesis, we follow two principles.
Firstly, a part hypothesis should cover as many as possible components of the corresponding ground-truth part.
Secondly, the component coverage of a part hypothesis should be conservative, meaning that
a hypothesis with missing components is preferred over that encompassing components across different
semantic parts.

\paragraph{\textbf{Grouping strategy}}
It is a non-trivial task to generate part hypotheses meeting the above requirements exactly.
It is very likely that there is not a single optimal criterion that
can be applied to generate hypotheses for any semantic part from a set of components.
For example, the many components of a car wheel can seemingly grouped based on a compactness criterion.
For bicycle chain, however, the tiny chain links are not compactly stacked at all,
for which size based grouping might be more appropriate.
Therefore, we design a grouping strategy encompassing three heuristic criteria,
which are intuitively interpretable and computationally efficient.
The grouping for each criterion is performed in a bottom-up fashion, based on a nested hierarchy.
After that, a hypotheses selection process is conducted to
ensure a conservative hypothesis coverage.
The quality of generated part hypotheses is evaluated in Section~\ref{subsec:hypo}.

\begin{figure}[t!]
  \centering
  \includegraphics[width=3.4in]{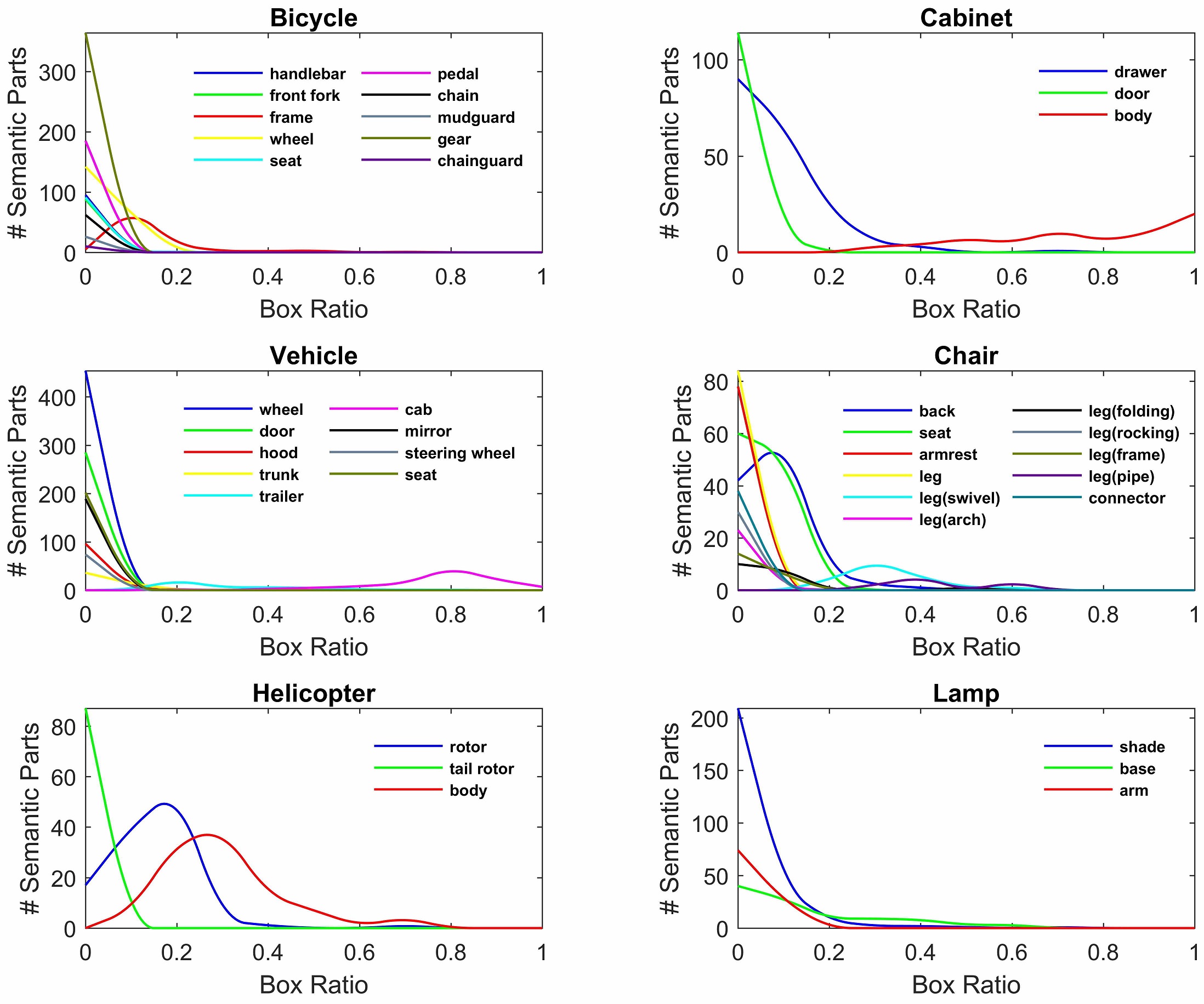}
  \caption{\new{The occupancy ratio of the bounding box of varying number of semantic parts over the entire model. The statistics are performed with our benchmark dataset.}}
  \label{fig:box_volume}
\end{figure}

Through a statistical analysis (Figure~\ref{fig:box_volume}),
%we can find that most of the semantic parts in the models are actually clustered within a relatively compact regions,such as wheel, steering wheel, etc.
we found that most semantic parts are spatially compact, such as a door or a wheel of a car.
We thus define a criterion called \emph{Center Distance}, denoted by $C_\text{center}(a,b)$,
to measure the compactness between two components $a$ and $b$.
It measures the distance between the barycenters of convex hull of two components,
and encourages grouping of components which are spatially close to each other.

The second criterion, sharing the similar intuition as center distance, imposes a stronger
test on compactness.
This is motivated by the fact that
the components in a functional part are typically tightly assembled.
The \emph{Geometric Contact} criterion, denoted as $C_\text{contact}(a,b)$,
prioritizes the grouping of components with large area of geometric contact.
Let $V_{a}$ and $V_{b}$ be the volume of component $a$ and $b$, respectively,
and $C_{ab}$ be the contact volume.
The criterion is defined as the maximum of the ratio between contact volume and component volume:
$$
C_\text{contact}(a,b) = \max\{ C_{ab} / V_a, C_{ab} / V_b \}.
$$
Here, the volume of a component can be computed by counting the voxels occupied by the component
in a global voxelization of the entire shape.
The contact volume between two components can be computed as the number of overlapping voxels of the two
components in the global voxelization.

\begin{figure}[t]
  \centering
  \includegraphics[width=3.4in]{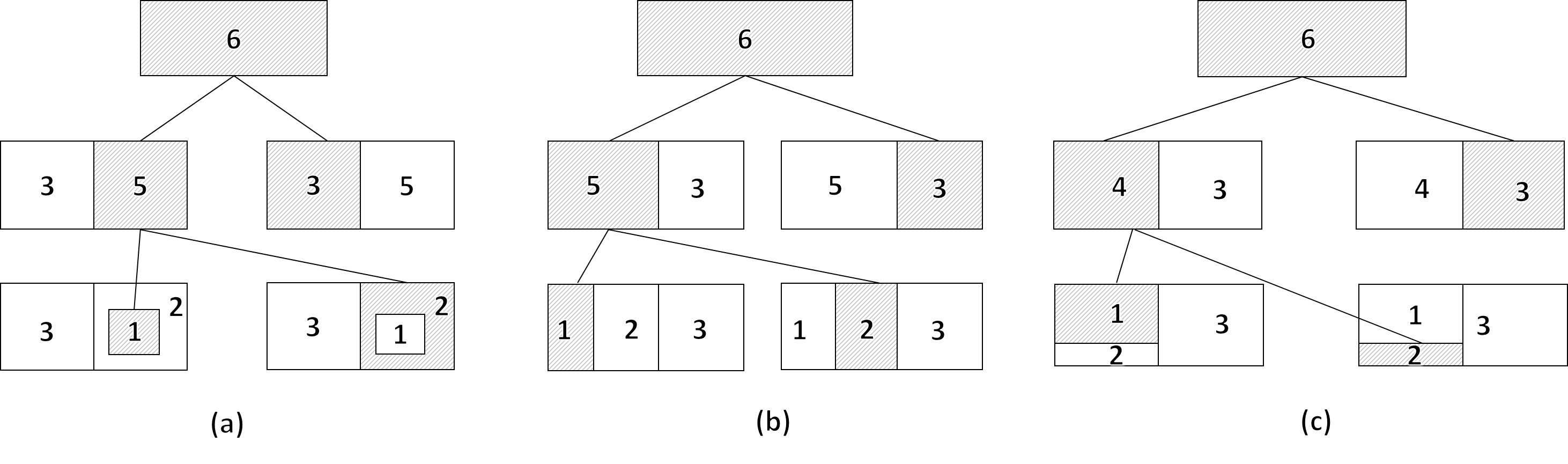}
  \caption{Hierarchical sampling of part hypotheses based on the three grouping criteria,
  center distance (a), group size (b) and geometric contact (c), respectively.}
  \label{fig:Rules}
\end{figure}

The semantic parts of a 3D model can be of arbitrary size,
ranging from a rear-view mirror to the entire cab for a car;
see the supplementary material.
Thus for each part category, we sample a set of candidate proposals with varying sizes, to avoid
missing the best one.
We design a third criterion \emph{Group Size}, denoted by $C_\text{size}(a,b)$,
as the occupancy rate of the joint volume of component $a$ and $b$ over the volume of the whole shape.
This criterion is used to control the grouping, sampling groups first in small size and then to large.

\paragraph{\textbf{Hierarchical hypothesis sampling}}
%As can be seen from the above definition, this is a very challenging task how to generate many high-quality part hypotheses for a 3D CAD shape.
%Moreover the sampling algorithm must meet the following requirements:
%1) Algorithm can capture all semantic parts scales, such as small scale rearview mirror, large scale tires.
%2) Abundant part hypotheses can be generated for each component.
%3) The sampling algorithm must be simple, fast and should not become a computational bottleneck of labelling.
%4) The number of samples should not be too large.
%
We employ a hierarchical aggregation algorithm to generate part hypotheses.
This is motivated by the fact that most off-the-shelf 3D models
are assembled with components in a hierarchical manner.
%moreover, the hierarchical aggregation algorithm is natural to meet condition 1,2.
%
Given a shape, the sampling process starts from the input set of components, and
groups in a greedy, bottom-up manner.
At each time, the pair of adjacent components with the smallest grouping criterion measure are
grouped into a new node.
The process is repeated until reaching the root of the hierarchy and performed for each criterion separately.
Figure~\ref{fig:Rules} illustrates the grouping process for each grouping criterion separately.
Nodes shaded in grey color in the hierarchies represent a sampled part hypothesis.
The top few groups in each hierarchy is then selected to form the candidate set for the given shape,
which will be discussed in the next.

\begin{figure}[!t]
  \centering
  \includegraphics[width=\linewidth]{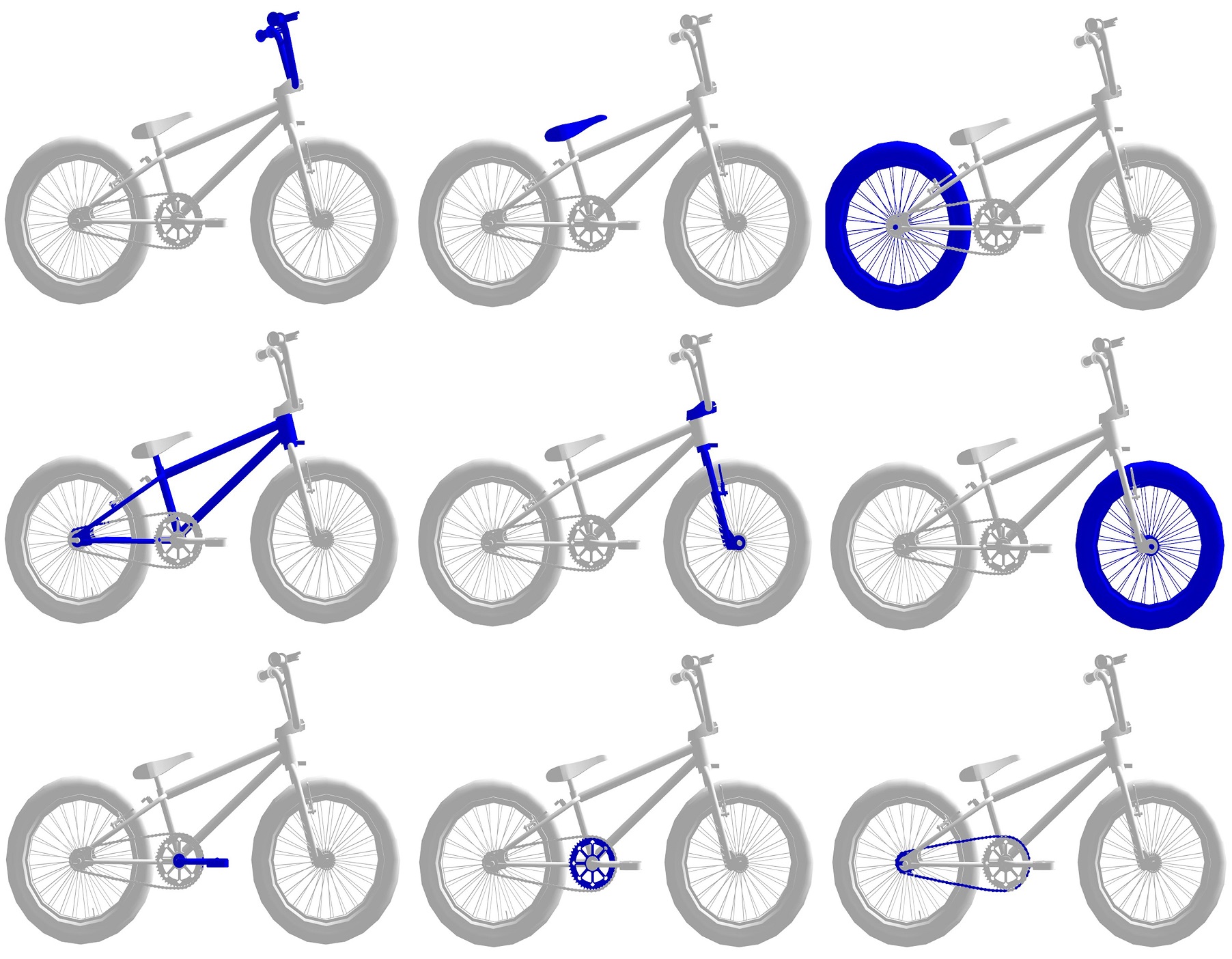}
  \caption{\new{Some samples of part hypothesis corresponding to different semantic labels of bicycle.}}
  \label{fig:Proposal_Bike}
\end{figure}

\paragraph{\textbf{Selection of part hypotheses}}
%For one grouping strategy, given a stock model with $n$ components,
%the sampled part hypotheses number is $2n-1$.
%As the number of components may be vary large, and to improve efficiency in subsequent process,
%we only keep a certain number of part hypotheses.
%
We first simply sort the hypotheses, corresponding to nodes in a hierarchy,
based on their grouping order.
Higher level nodes imply larger coverage, while lower ones correspond to smaller regions.
To prevent the selection from overly favoring hypotheses large coverage,
we introduce random factors into the selection process, in a
similar spirit to~\cite{Carreira_PAMI12,Manen2014Prime,Van2011Segmentation}.
In particular, the initial sorting is perturbed by multiplying the sorted indices with a random number in $(0,1)$,
and then resorting based on the resulting numbers. Finally, the top $H$ hypotheses are selected for each hierarchy,
thus yielding $3H$ hypotheses in total.
In Figure~\ref{fig:Proposal_Bike},
we visualize a few part hypotheses corresponding to some semantic parts for a bicycle model.

\begin{figure}[!t]
  \centering
  \includegraphics[width=3.4in]{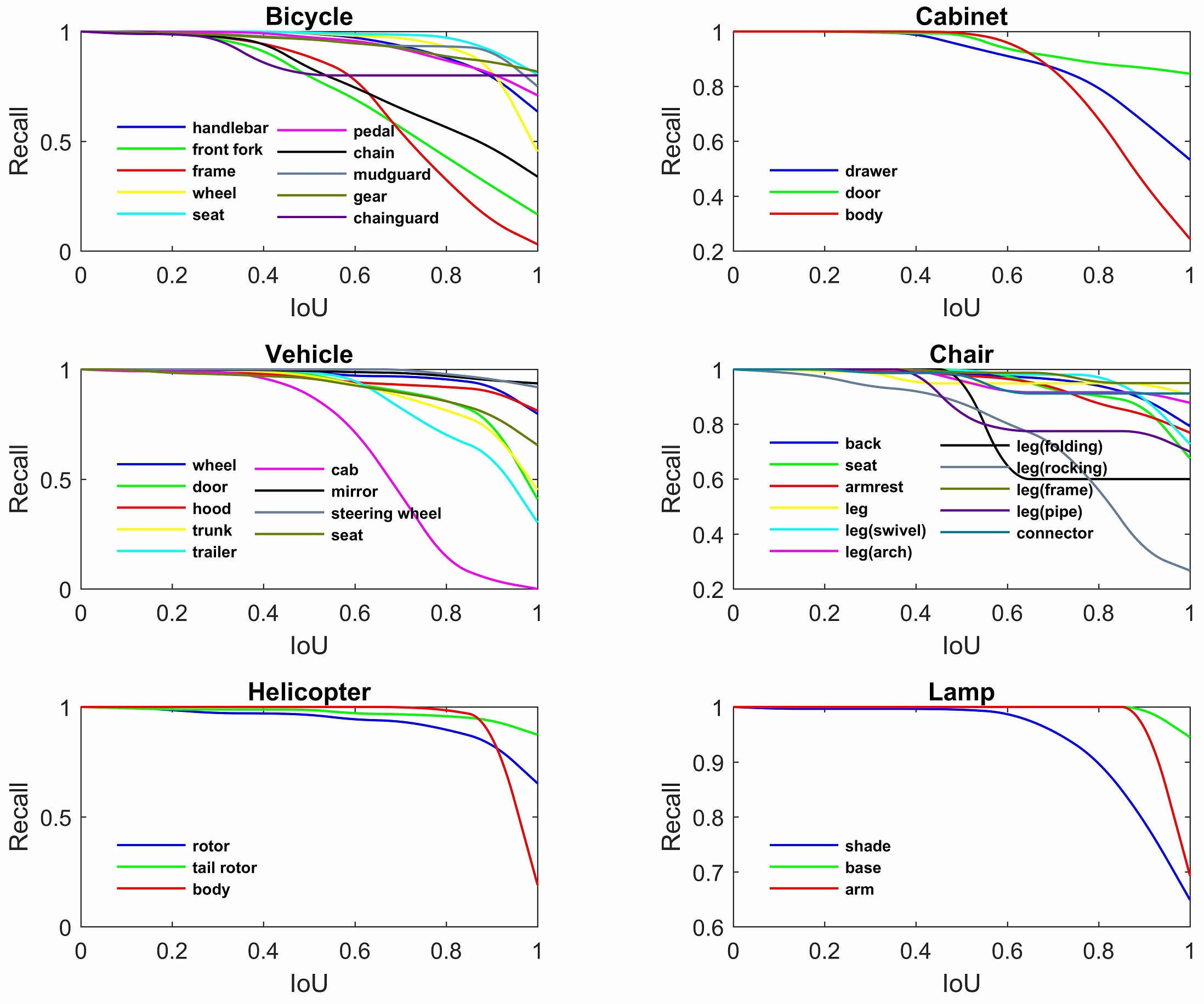}
  \caption{Performance (recall rate over IoU) of part hypothesis generation in all object/scene categories.}
  \label{fig:IoURecall}
\end{figure}

\kx{
The Intersection of Union (IoU) based recall (the recall rate for a given IoU threshold w.r.t. ground-truth) of the hypothesis selection is given in Figure~\ref{fig:IoURecall}.
It shows that our hierarchical sampling and selection method is quite effective in
capturing the potential semantic parts, even for complicated structures such as chairs, bicycles
and helicopters.
Note that although the recall rates drop significantly around the IoU threshold of $0.6$ for several categories, it does not hurt the performance since the recall rates for IoU of $0.6$ are already high enough for the following CRF-based labeling algorithm to perform well. Results in Figure~\ref{fig:Performance_vs_num} show that the labeling accuracy is stable with the number of sampled part hypotheses and $1000$ proposals (our default choice) are sufficient for all categories.
}

\paragraph{\textbf{Remarks on design choice}}
\kx{For part hypothesis generation, we opted for combinatorial search with a hierarchical guidance rather than a learning based approach. This is because the significantly varying number and size of components within a semantic part make it extremely difficult for, e.g., a CNN model, to capture the shape geometry with a fixed input resolution. Taking the part hypotheses in a bicycle model (Figure~\ref{fig:Proposal_Bike}) for example, the component size and count differ greatly from part to part. On the other hand, a proper resolution of data representation for CNN is unknown before the part hypothesis is extracted -- a chicken-and-egg problem! We have implemented and compared with a CNN-based method as a baseline, demonstrating clear advantage of our approach (Section~\ref{subsec:label} and~\ref{subsec:hypo}).}

\begin{figure*}[ht]
  \centering
  \includegraphics[width=0.9\textwidth]{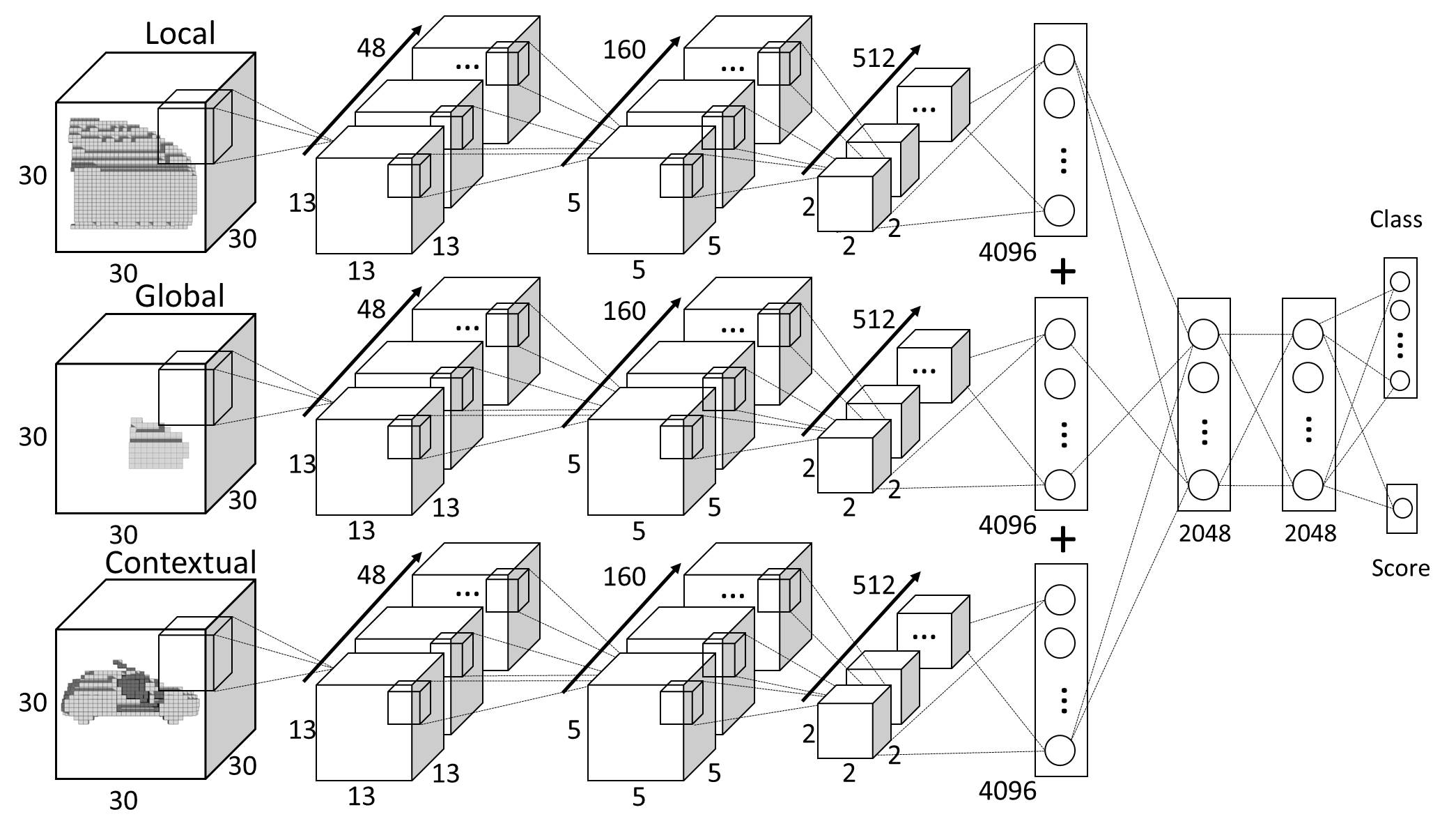}
  \caption{The architecture of our multi-scale Convolutional Neural Networks (CNNs) for part hypothesis classification and confidence regression.}
  \label{fig:CNN}
\end{figure*}

\subsection{Classifying and Scoring of Part hypotheses}
\label{sec:ranking}
We train a neural network to classify a part hypothesis and produce a confidence score for it
representing the confidence of the hypothesis being an independent semantic part.
To achieve that,
we first build a training dataset of multi-component 3D models
with component-wise labels.
We then design a multi-scale Convolutional Neural Networks (CNNs), which learns
feature representation capturing not only local part geometry but also global context.

\paragraph{\textbf{Training data}}
The multi-component 3D models used for training are collected from both ShapeNet~\cite{ShapeNet2015} and 3D warehouse~\cite{Tri3Dwarehouse}.
The data comes from the training part of our multi-component labeling benchmark (see Section~\ref{subsec:benchmark}).
Each model in the training set has a component-wise labeling, based on the semantic labels defined
with WordNet. \kx{An overview of the human-labeled models from the dataset can be found in the supplementary material.}
The training set contains eight object categories and two scene categories on which the statistics are detailed in Table~\ref{tab:LabellingAssessment}.

%including {vehicle} ($109$), bicycle ($98$), {chair} ($114$), {cabinet} ($82$), {plane} ($95$), {lamp} ($100$),
%{motor} ($109$), and {helicopter} ($105$))
%as well as two scene categories: living room ($102$) and office ($102$).
%The numbers in parentheses are model count.
%There are $1016$ models in total. [PLEASE CONFIRM!]}

%\subsection{Multi-scale CNNs}
\paragraph{\textbf{Input data representation}}
To train a CNN model for part hypothesis labeling,
we opt for volumetric representation of part hypotheses as input, similar to~\cite{Wu_CVPR15}.
%We desire to describe the whole structure of part hypothesis which is assembled by multiple components, not just appearance.
%Due to computational cost, volume resolution should not be too large. However, for the description of part hypotheses, a large resolution is required
%to ensure the local geometric detail feature to be valid.
%Simultaneously, we desire the description to preserve the contextual information.
To achieve a multi-scale feature learning,
we represent each hypothesis in three scales including
a \emph{local} scale based on a voxelization of its bounding box,
a \emph{global} scale, which takes the volume of the bounding box of the entire shape and
contributes two channels.
One channel encodes the volume occupancy of the part hypothesis itself
and the other accounts for the context based on the occupancy of the remaining parts.
%\textbf{\textsl{Local}}: Normalization of candidate to preserve the local detail;
%\textbf{\textsl{Global}}: Keep the relative position of candidate with the remaining components;
%\textbf{\textsl{Remaining}}: The remaining components to encode the contextual information.
For each scale, the volume resolution is fixed to $30\times30\times30$.
\wang{
To avoid the global alignment among all shapes, we opt for training with many possible orientations of each shape. In practice, we use the up-right orientation of each shape and enumerate its four canonical orientations (Manhattan frames).
}

\paragraph{\textbf{Data augmentation for balanced training}}
\new{
The hierarchical grouping of part hypotheses could
make the training data unbalanced: Insufficient data is sampled for
semantic categories containing small number of components (e.g., rear-view mirrors of cars).
This will make our CNN model inadequately trained for these categories.
To cope with this issue,
we opt to synthesize more training data, for the categories with insufficient instances,
based on the ground-truth of semantic parts in the training data.
Specifically, we pursue two ways for data augmentation.
\begin{enumerate}
  \item \textsl{Component deletion}. Given a ground-truth semantic part, we randomly delete a few components and use the incomplete part as a training example. We typically remove up to $30\%$ components.
  \item \textsl{Component insertion}. Given a ground-truth semantic part, we randomly insert a few components from the neighboring parts to form a training example. We stipulate that the newly added components do not exceed $30\%$ of original ones.
\end{enumerate}
}

\paragraph{\textbf{Ground-truth labels and scores}}
We next compute a ground-truth part label and confidence score for each training part hypothesis,
used for training our network for both label prediction and score regression.
For a given part hypothesis, if its components labeled with a certain category
occupy over $70\%$ of the global voxelization of the entire shape, it is treated as a positive example
for that category, and negative otherwise.
For each hypothesis, we first compute its 3D Intersection of Union (IoU)
against each ground-truth semantic part of the shape, in a global voxelization of $200\times200\times200$.
The highest IoU is set as its confidence score.
This score measures the confidence of a hypothesis being an independent semantic part,
which will be utilized in the final label inference in Section~\ref{sec:labelling}.

\paragraph{\textbf{Network architecture}}
We design a multi-scale Convolutional Neural Networks (CNNs).
The architecture of our network is given in Figure~\ref{fig:CNN}.
The network has three towers, taking the inputs corresponding to
the local and the two global channels mentioned above.
We refer to these towers as \emph{local}, \emph{global} and \emph{contextual}, respectively;
see Figure~\ref{fig:CNN}.
The feature maps output by the three towers are concatenated into one feature vector,
and then fed into a few fully connected layers, yielding a $2048$ feature vector.
The final fully connected layer predicts
a label and regresses a score for the input part hypothesis.
In particular, our network produces a probability distribution over $K+1$ part labels, $p=(p_0,\ldots,p_K)$,
with label $0$ being null label.
An unrecognized part is assigned with a null label.
The second output is confidence score $r$.
\new{
We use a joint loss $L$ for both hypothesis classification and score regression:
\begin{eqnarray}
\begin{aligned}
  L(p,r,c,s)= L_\text{cls}(p,c) + L_\text{reg}(r,s),
\label{eq:multi_task_loss}
\end{aligned}
\end{eqnarray}
where $L_\text{cls}(p,c)=-\log p_{c}$ is cross-entropy loss for label $c$.
and $L_\text{reg} $ is smooth $L_1$ loss.
}

\begin{table*}[ht]
\caption{\wang{Accuracy of grouping and labeling (average Intersection of Union, in percentage) on our benchmark dataset MCL.
Row 1: The average number of components for each category.
Row 2: The number of annotated semantic labels for each category.
Row 3: The maximum numbers of part hypotheses for each model used for final labeling inference.
Row 4: Training / testing split (number of models) of our dataset.
Row 5: The number of training examples (hypotheses) for each category.
Row 6-20: Average IoU of baseline methods, state-of-the-art methods and our method in different settings.}}
  \centering
    \begin{tabular*}{\textwidth}{lp{0.20\textwidth}<{\centering}|ccccccccccc}
    \hline
    \multicolumn{2}{c|}{Rows} & Vehicle & Bicycle & Chair & Cabinet & Plane & Lamp  & Motor & Helicopter & Living room & Office \\
    \hline \hline
    1&\# Avg. components & 649  & 572 & 31 & 53 & 111 & 17 &  188 &  178  &  197  & 276\\
    2&\# Semantic labels & 9 & 10 & 11 & 3 & 10 & 3 & 9 & 3 & 8 & 7\\
    3&\# Part hypotheses ($\leqslant$) & 1000 & 1000 & 200 & 200 & 1000 & 200 & 1000 & 1000 & 200 & 200 \\
    4&\# Train  / \# Test  & 26 / 83 & 30 / 68  & 33 / 81  & 25 / 57 & 17 / 78 & 30 / 70  & 22 / 87  & 21 / 84 &  30 / 72 &  30 / 72\\
    5&\# Training hypo. &23787 & 28947 & 2149 & 4666 & 4812 & 1662 & 12800 & 8263 & 7090 & 14475\\
    \hline \hline
    6&{\small Baseline (Random Forest)} & 54.7 & 58.9 & 62.4 & 65.9 & 53.5 & 63.3 & 65.9 & 52.8 & 47.7 & 68.5\\
    7&{\small Baseline (CNN Classifier)} &48.9 & 63.8  & 70.75  & 63.3 & 68.9 & 81.2 & 67.4 & 78.5 &51.2 &63.9\\
    8&{\small Baseline (CNN Hypo. Gen.)} &56.3 &51.9 &68.5 &45.7 &58.5 &71.1 &53.1 &72.2 &58.6 &69.1\\
    \hline \hline
    9&PointNet~\citep{Su_CVPR17} & 24.3 & 30.6 & 68.6 & 21.0 & 47.2 & 46.3 & 35.8 & 32.6 & - & - \\
    10&PointNet++~\citep{qi_2017} & 51.7 & 53.8 & 69.3 & 62.0 & 53.9 & 79.8 & 62.2 & 79.3 & - & - \\
    11&Guo et al.~\shortcite{Guo_TOG15} & 27.1 & 25.2 & 34.2 & 68.8 & 38.6 & 79.1 & 41.6 & 80.1 & 33.7 & 28.5\\
    12&Yi et al.~\shortcite{Yi_SG17} & 65.2 & 63.0 & 61.9 & 70.6 & 59.3 & 82.2 & 67.5 & 78.9 & 56.6 & 68.6\\
    \hline \hline
    13&Ours (w/o score) & 71.5 & 66.8 & 72.5 & 76.5 & 71.4 & 87.6 & 70.7 & 81.2 & 63.3 & 60.1\\
    14&Ours (local only) & 50.4	&52.4	&60.4	&68.6	&61.3	&73.5	&60.4	&78.5	&62.7	&54.8\\
    15&Ours (local+global) & 69.2	&67.3	&68.6	&75.4	&69.1	&79.2	&67.2	&82.6	&\textbf{68.3}	&\textbf{76.4}\\
    16&Ours ($n=1$) & 52.0 & 43.2 & 63.5 & 62.0 & 47.6 & 76.5 & 41.7 & 42.4 & 54.6 & 70.7\\
    17&Ours ($n=3$) & 56.5 & 49.9 & 67.0 & 66.6 & 55.4 & 84.0 & 51.7 & 43.4 & 63.1 & 70.1\\
    18&Ours ($n=5$) & 59.3 & 54.9 & 70.5 & 69.6 & 59.8 & 86.3 & 55.3 & 50.7 & 64.7 & 68.9\\
    19&Ours ($n=10$) & 62.0 & 61.9 & 72.6 & 74.1 & 68.6 & 86.9 & 62.4 & 75.6 & 66.6 & 66.1\\	
    20&Ours ($all$) & \textbf{73.7} & \textbf{68.1} & \textbf{74.3} & \textbf{78.7} & \textbf{76.5} & \textbf{88.3} & \textbf{71.7} & \textbf{83.3} & 66.1 & 65.4\\
    \hline
  \end{tabular*}
\label{tab:LabellingAssessment}
\end{table*}

\if 0
\begin{table*}[ht]
  \centering
    \begin{tabular*}{\textwidth}{lp{0.18\textwidth}<{\centering}|ccccccccccc}
    \hline
    \multicolumn{2}{c|}{Rows} & Vehicle & Bicycle & Chair & Cabinet & Plane & Lamp  & Motor & Helicopter & Living room & Office \\
    \hline \hline
    1& p-value & 0.9949  & 0.9180 & 0.9975 & 0.9772 & 0.9973 & 0.9370 &  0.9941 &  0.9839  &  0.9888  & 0.0006\\
\hline
  \end{tabular*}
\label{tab:xxx}
\end{table*}
\fi

\subsection{Composite Inference and Labeling}
\label{sec:labelling}
\begin{figure}[t]
  \centering
  \includegraphics[width=\linewidth]{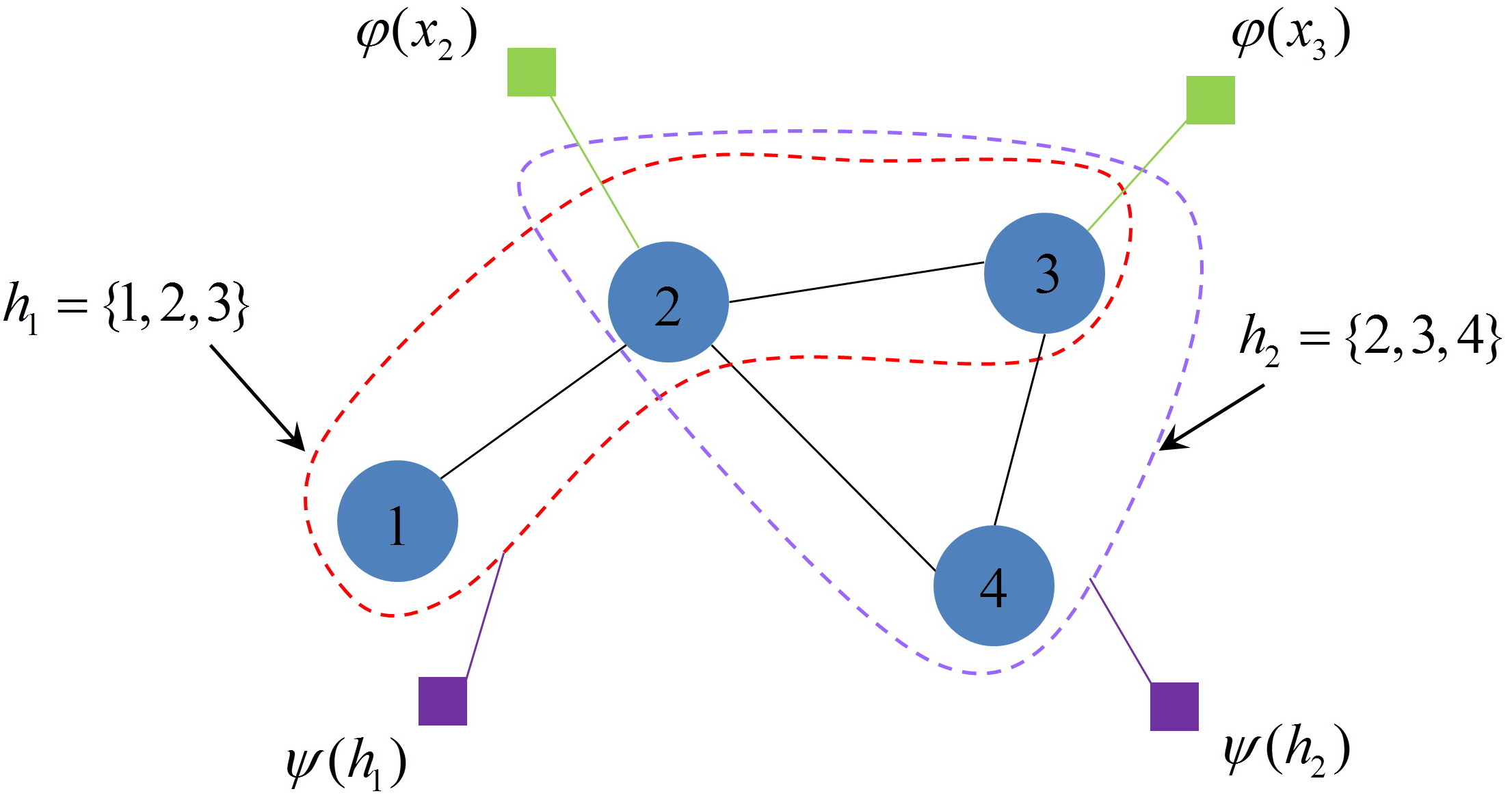}
  \caption{Illustration of our higher-order CRF.
$\varphi(x_{i})$ is unary potential.
$\psi(\mathbf{x}_{h})$ is higher order consistency potential, which favours all components belonging to a part hypothesis taking the same label.}
  \label{fig:CRF}
\end{figure}

\new{
Given the confidence scores of a sampled part hypothesis,
the final stage of our method is to infer an optimal label assignment for each component.
Given a multi-component 3D model, denoted by $M$, which comprises a set of components $\mathcal{C}$.
Each component $c \in \mathcal{C}$ is associated with a random variable $x_c \in X$
which takes a value from the part label set $\mathcal{L}=\{l_1,\ldots,l_K\}$.
Let $\mathcal{H}$ denote the set of all part hypotheses.
A part hypothesis $h \in \mathcal{H}$ is denoted by a set of components,
$h \coloneqq \{c^h_i\}_i \subset \mathcal{C}$, and its labeling is represented
a vector of random variables $\mathbf{x}_{h} = (x^h_i)_i$, with $x^h_i \in X$ being the label
assignment for component $c^h_i$.
A possible label assignment to all components, denoted by $X$, is called a \textsl{labeling} for model $M$.
\if 0
Given a multi-component model, denoted by $M$, which comprises a set of components $\mathcal{C}=\{c_i\}_{i=1}^C$.
% ,where $N$ is the total number of components.
Each component $c_i \in \mathcal{C}$ is associated with a random variable $x_{i} \in X$
which takes a value from the part label set $\mathcal{L}=\{l_i\}_{i=1}^L$.
%category set $\mathcal{L}=\{l_{1},\cdots,l_{K}\}$.
Let $\mathcal{H}=\{h_i\}_{i=1}^H$ denote the set of all part hypotheses.
A part hypothesis $h$ is denoted by a set of components,
$h_i \coloneqq \{c^i_j\} \subset \mathcal{C}$, and its labeling is represented
a vector of random variables $\mathbf{x}_{i} = (x^i_1,\ldots,x^i_C)$
Every possible assignments of part categories to the random variables $X$ will be called a \textsl{labeling} which takes values from the $L=\mathcal{L}^N$.
\fi

We construct a higher-order Conditional Random Fields (CRFs),
to find the optimal labeling for all components,
based on the part hypothesis analysis from the previous steps:
%describe the properties of components and their $assembly$ relationship in the CAD model,
\begin{eqnarray}
%\begin{equation}
\begin{aligned}
  E(L)=\sum_{c \in \mathcal{C}} {\varphi(x_{c})}+  \lambda\sum_{h \in \mathcal{H}}{ \psi(\mathbf{x}_{h})},
\label{eq:CRF}
\end{aligned}
%\end{equation}
\end{eqnarray}
where the first term is the unary potential for each component and
the second term is the higher order consistency potential defined with each hypothesis.
The parameter $\lambda$ is used to tune the importance of the two terms.
We set $\lambda=0.1$ in all our experiments.
The CRF-based labeling is illustrated in Figure~\ref{fig:CRF},

\paragraph{\textbf{Unary potential}}
%$For component $i \in \mathcal{C}$,
Suppose $\mathcal{H}^{c}=\{h^c_i\}_{i=1}^{H^c}$ be the set of part hypotheses containing component $c$.
The unary potential $\varphi(x_{c})$ is defined as:
\begin{eqnarray}
%\begin{equation}
\begin{aligned}
  \varphi(x_{c})= -\log P(x_{c}=l_k),
\label{eq:CRF_unary}
\end{aligned}
%\end{equation}
\end{eqnarray}
where $P(x_c=l_k)$ is the probability of $x_c$ taking the label $l_k$,
and is defined as:
\begin{eqnarray}
\begin{aligned}
  P(x_c=l_k)= \frac{{\sum_{i=1}^{K^c}}{e^{w_i^c  s_i^c  p(l_k|h^c_i)}}}{{\sum_{k=1}^K} {\sum_{j=1}^{K^c}}{e^{w_j^c s_j^c p(l_k|h_j^c)}}},
\label{eq:CRF_unary_P}
\end{aligned}
\end{eqnarray}
where $p(l_k|h_i^c)$ is the classification probability of hypothesis $h_i^c$ against label $l_k$,
output of our hypothesis classification network.
$K^c \leqslant H^c$ is the top number of part hypotheses selected for computing the probability,
based on the regressed confidence score for $c$.
$s_i^c$ is the confidence score for $h_i^c$, regressed by our network.
$w_i^c$ is a weight computed as the ratio between the volume of component $c$
and that of hypothesis $h_i^c$.

\paragraph{\textbf{Higher order consistency potential}}
The goal of our CRF-based labeling is to resolve the inconsistency between different
part hypotheses and compute a consistent component-wise labeling,
resulting in a non-overlapping partition of all components.
To this end, we design a higher order consistency potential~\cite{Kohli2008Graph,Park2012On},
based on the \emph{label purity} of part hypotheses:
\begin{eqnarray}
%\begin{equation}
\begin{aligned}
   \psi(\mathbf{x}_{h})=
   \left\{
    \begin{array}{rcl}
           N(\mathbf{x}_{h})\frac{1}{\eta}\gamma_\text{max}, &    & \text{if  $N(\mathbf{x}_{h}) \leq \eta$}\\
           \gamma_\text{max},    &    & \text{otherwise}
	\end{array}
   \right.
\label{eq:CRF_consistency}
\end{aligned}
%\end{equation}
\end{eqnarray}
where $N(\mathbf{x}_{h})=\min_{k}\{C^h-n_{k}(\mathbf{x}_{h})\}$.
$C^h$ is the number of components constituting part hypothesis $h$.
$n_{k}(\mathbf{x}_{h})$ counts the number of random variables corresponding to hypothesis $h$
which takes label $k$.
$\eta$ is the truncation parameter which controls the rigidity of the higher order consistency potential,
and is set to $0.2*C^h$ in all our experiments.
This means that up to $20\%$ of $h$'s components can take an arbitrary label.
$\gamma_{max}=e^{-G(h)/C^h}$,
with $G(h)$ being the label purity of a part hypothesis.
The purity can be computed as the entropy of the classification probability output of our network:
\begin{eqnarray}
%\begin{equation}
\begin{aligned}
  G(h)=-\sum_{k=1}^K p(l_k|h) \log p(l_k|h).
\label{eq:CRF_consistency_G}
\end{aligned}
%\end{equation}
\end{eqnarray}
where $p(l_k|h)$ is the classification probability of hypothesis $h$
against label $l_k$ output, again, by our network. } %%%%%%%%%%%%%%end

The consistency potential encourages components
belonging to one part hypothesis to take the same label.
However, it does not impose a hard constraint on label consistency by allowing
a portion of the components within a part hypothesis to be labeled freely.
This is achieved by the linear truncated cost over the number of inconsistent labels.
This mechanism enables the components within a part hypothesis to be assigned to different labels,
so that an optimal label assignment to all components could be found through compromising among
all part hypotheses.
%Finally,
%we solve Equation~\ref{eq:CRF} by using the expansion and swap move algorithms~\cite{Kohli2008Graph}.
%The optimization performs effectively,
%producing the final consistent emantic labelling results.
The objective in Equation (\ref{eq:CRF}) can be efficiently optimized
with the alpha-beta moving algorithm~\cite{Kohli2008Graph}.

An alternative approach to CRF-based labeling would be formulating it as a deep learning model. The combination of CRF and deep neural networks has shown promising results on semantic segmentation of 2D images~\cite{zheng2015}. In the context of 3D component grouping and labeling, however, training such a deep model requires a large amount of relational data between different components, which is highly laborious. We believe our solution achieves a good balance between model generality and complexity.

\section{Results and Evaluations}
\label{sec:result}
%\todo{We present both qualitative and quantitative evaluations for each stage of our approach.}

\subsection{Multi-Component Labeling benchmark}
\label{subsec:benchmark}
To facilitate quantitative evaluation, we construct the first benchmark dataset with
human-annotated, component-wise labels, named multi-component labeling benchmark, or \emph{MCL benchmark} for short.
The multi-component 3D models are collected from ShapeNet~\cite{ShapeNet2015} and 3D warehouse~\cite{Tri3Dwarehouse},
in which most 3D models are in the form of multi-component assembly.
We manually annotate each model in our dataset
by assigning a semantic category to each component, using our interactive annotation tool.
\kx{The annotation tool is elaborated in the supplementary material.
The semantic part categories are defined based on WordNet, which are summarized with an overview of the benchmark dataset in the supplementary material. Some statistics of the dataset are also given therein.}

\if 0
\begin{figure}[ht]
  \centering
  \includegraphics[width=3.4in]{Images/our_dataset_components_stat}
  \caption{\new{Statistics on components for our multi-component labeling (MCL) benchmark dataset.}}
  \label{fig:mclstats}
\end{figure}

\begin{figure*}[ht]
  \centering
  \includegraphics[width=\textwidth]{Images/Dataset}
  \caption{\new{An overview of our multi-component labeling (MCL) benchmark dataset. More can be found in the supplementary material.}}
  \label{fig:Dataset}
\end{figure*}
\fi

Row 1 and 2 of Table~\ref{tab:LabellingAssessment} provides a summary and detailed statistics
about our MCL benchmark dataset.
For each category, about $20\%\sim30\%$ models are used for training,
and the remaining for testing.
Such a training/testing split is fixed all subsequent experiments.
%The benchmark dataset comes with a \kx{training ($70\%$) / validation ($10\%$) / test ($20\%$) split}.
A few metrics on segmentation accuracy are defined to support quantitative evaluation
of component labeling; see the following subsections for details.
In the supplementary material, we provide an overview of the benchmark.
We believe this benchmark would benefit more future research
on component-wise shape analysis and data-driven shape modelling~\cite{Sung2017,li_sig17}.

\subsection{Labeling performance}
\label{subsec:label}
\new{
We evaluate our semantic labeling based on our MCL benchmark.
The performance is measured by average Intersection of Union (avg IoU).
The results, reported in the last row of Table~\ref{tab:LabellingAssessment},
%Some semantic labelling results on raw 3D CAD shape generated by our approach are shown in Figure~\ref{fig:MoreResults}.
show that our approach achieves the best performance.
In Figure~\ref{fig:MoreResults}, we show visually the labeling results.
Our approach is able to produce robust labeling for fine-grained components with
complex structure and severe self-occlusions.

We also test our method on the INRIA GAMMA 3D Mesh Database~\citep{GAMMA}, which is a large collection of human created 3D models. Our method, trained on the our MCL benchmark dataset, is applied to INRIA GAMMA database.
Figure~\ref{fig:GAMMA_database} presents some labeling results on a few sample models, produced by our method.
More results can be found in the supplementary material.
%https://www.rocq.inria.fr/gamma/gamma/download/download.php

\begin{figure}[t]
  \centering
  \includegraphics[width=3.4in]{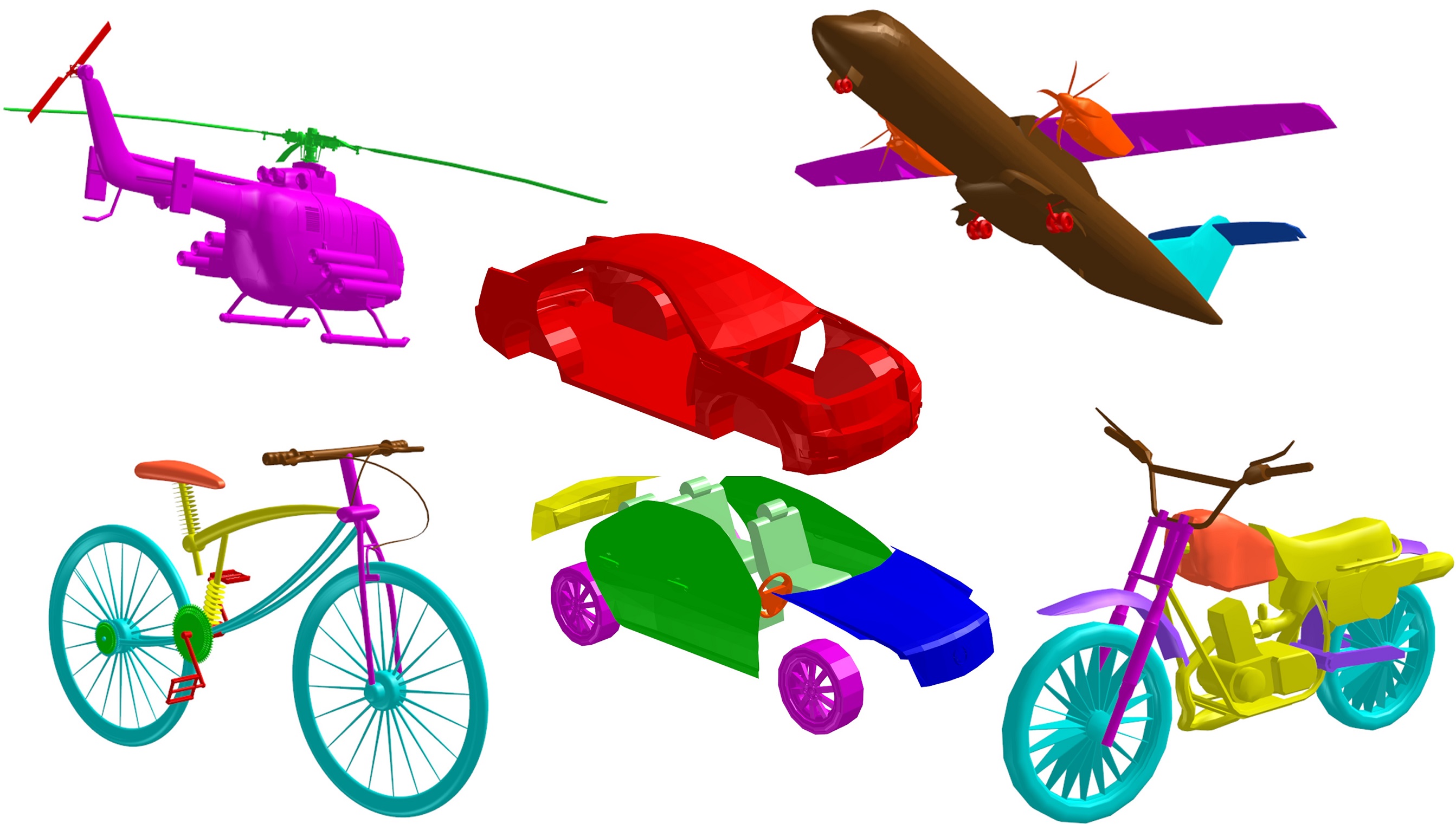}
  \caption{\new{Labeling results on five sample models from the INRIA GAMMA models produced by our method.}}
  \label{fig:GAMMA_database}
\end{figure}

\paragraph{\textbf{Comparison with baseline (random forest)}}
To verify the effectiveness of our part hypothesis based analysis and multi-scale CNN based labeling,
we implement a baseline using conventional method, i.e., hand-crafted features plus random forest classification. Specifically, we extract features, including light field descriptor~\cite{Chen2010On}, spherical harmonic descriptor~\cite{Kazhdan2003Rotation}, volume ratio and bounding box diameter, for each component,
and feed them into a random forest classifier for component classification.
We used the default parameter settings of the standard MATLAB toolbox for random forest,
with the number of trees being $500$.
The comparison is shown in Table~\ref{tab:LabellingAssessment} (row 6).
Our mid-level, part hypothesis analysis (the last row) significantly outperforms this alternative.

\paragraph{\textbf{Comparison with baseline (CNN-based classification)}}
The second baseline we compare to is a direct CNN-based component classification, without
part hypothesis based analysis.
Taking labeled components as training samples,
we learn the network with the same architecture in Figure~\ref{fig:CNN},
except that only a classification loss, $L_{cls}$, in Equation (\ref{eq:multi_task_loss}) is employed.
The performance is reported in Table~\ref{tab:LabellingAssessment} (row 7).
Our hypothesis-level analysis (the last row) achieves much higher
labeling accuracy than component-level analysis, due to the fact that part hypotheses
capture richer semantic information than individual components.

\paragraph{\textbf{Comparison with baseline (CNN-based hypothesis generation)}}
\label{subsec:base}
\kx{
To demonstrate the difficulty of part hypothesis generation from fine-grained components with drastically varying numbers and sizes, we implement a CNN-based hypothesis generation through extending Fast RCNN~\cite{girshick2015} to 3D volumetric representation.
The network architecture and its detailed explanation can be found in the supplemental material.
Taking the volumetric representation of a shape as input, the network is trained to predict at each voxel
a 3D box representing part hypothesis.
This is followed by another network for joint classification and refinement of the hypothesis regions.
The training data utilize the ground-truth parts in our MCL dataset after voxelization.
The results shown in Table~\ref{tab:LabellingAssessment} (row 8) are inferior to those of our method.
The main reason is that the significant scale variation of components makes it difficult for volumetric representation to characterize their shape and structure. This justifies our design choice of hierarchical search for part hypothesis generation.

%Then, a classification and a refinement branch are devised for classifying and refining the hypotheses.
%In classification branch, each hypothesis is transformed into a fixed-size feature vector by max-pooling and fully connected layers.
%It outputs a discrete probability distribution (per proposal), $p=(p_0,...,p_K)$, over $K$+1 categories.
%Refinement branch takes each component in current proposal as input,
%and outputs a probability distribute, $b=(b_0,b_1)$, over two categories (is adopted by current proposal vs. not).

%For each occupancy voxel location, we will predict $N$ candidate proposals.
%Each of the proposal corresponds to one of the $N$ boxes with with various sizes.
%In our case,based on statistics of semantic parts sizes in our dataset, we define a set of $N$=30 boxes.
%Note that, our proposal is not a regular cube region, but the region that related components covered in this box.

%Each training proposal is labeled with a ground-truth semantic class $u$,
%and each component in proposal has a binary label $v$, which represents whether the component should be adopted by the proposal.
%We use a multi-task loss $L$ on each labeled proposal to jointly train
%for classification and refinement:
%\begin{eqnarray}
%\begin{aligned}
%  L(p,u,b,v)= L_\text{cls}(p,u) + \sum_{c \in h} L_\text{mask}(b,v),
%\label{eq:multi_task_loss}
%\end{aligned}
%\end{eqnarray}
%where $L_\text{cls}(p,u)=-\log p_{u}$ is cross-entropy loss for label $u$.
%And for each component $c$ in current proposal $h$, $L_\text{mask}=-\log c_{v}$ is log loss over two categories (is adopted by proposal vs. not).
}

\paragraph{\textbf{Comparison to state-of-the-art methods}}
We compare our approach with the methods in~\citep{Yi_SG17} and~\cite{Guo_TOG15},
both of which adopt multiple traditional features as inputs to train neural networks.
For the shapes in our dataset,
we compute both face-level and component-level geometric features, based on the original implementation
of the two works.
%Face-level features include spin images, shape context, distance distribution,
%local PCA, local point position variance, curvature, point position and normal.
%Component-level features include light field descriptor, center-of-mass, bounding box diameter,
%approximate surface area, and local frame of PCA.
Details on the features can be found in the two original papers respectively.
Note, however, the work~\citep{Yi_SG17} is able to produce hierarchical labeling while our method
is not designed for this goal.
To make the two methods comparable, we compare our labels to those of only leaf nodes produced by~\citep{Yi_SG17}.

\wang{
Our method is also compared with PointNet~\cite{Su_CVPR17} and PointNet++~\cite{qi_2017},
two state-of-the-art deep learning based methods for semantic labeling of point clouds.
%a state-of-the-art deep learning based method for semantic labeling of point clouds.
We apply these methods by sampling the surface of the test shapes,
while keeping the semantic labeling, resulting in about $10K$ points for each shape.
To ensure a good performance of the two methods on our dataset and a fair comparison, we used their models pre-trained on ShapeNet and fine-tuned them on our training dataset.
}

We report per-category IoU percentage of these \wang{four} methods on our benchmark dataset, see Table~\ref{tab:LabellingAssessment}.
The results demonstrate the significant advantage of our part hypothesis analysis approach,
with consistently more accurate labeling.
In particular, our method significantly outperforms~\cite{Yi_SG17} on all categories and is comparable on `office'.
The significance is high (p-value $>0.98$) for models with severe self-occlusions such as
vehicles, cabinets, motors, etc., and moderately high (p-value $>0.92$) for category `bicycle' and `lamp'.
Another notable observation is that, all the alternative methods, especially PointNet \wang{and PointNet++},
find a hard time in dealing with scene models.
Scenes typically have more complicated structures due to the loose spatial coupling between objects.
Our method, on the other hand, is able handle structures in various scales and forms, ranging
from individual objects to compound scenes.
%As shown in Figure~\ref{fig:CompVsPoint}, the reason for 'failure' of PointNet is probably due to that the point clouds seriously lost the local detail and structural context with respect to the 3D CAD shape, while our approach effectively preserve this information by 'part hypotheses'.
}

\if 0
\begin{figure}[!t]
  \centering
  \includegraphics[width=3.0in]{Images/CompVsPoint}
  \caption{\new{A bike represented by raw 3D CAD shape comprised $523$ components (left), and represented by $10000$ point clouds (right).}}
  \label{fig:CompVsPoint}
\end{figure}
\fi

%\paragraph{Comparison with State-of-the-art method  ~\cite{Su_CVPR17}.}

\if 0
\begin{figure*}[ht]
  \centering
  \includegraphics[width=\textwidth]{Images/baseline_fast_rcnn}
  \caption{\wxg{Baseline 3D proposal generation network. The network takes a complete shape as input.
Then multi-scale boxes are applied to produce candidate regions on different scales and align corresponding feature maps.
Classification and Refinement branches are responsible for classifying and refining candidate regions respectively.
In which, classification branch outputs a discrete probability distribute for each candidate region over $K$+1 categories.
Refinement branch takes each component in current proposal as input,
and outputs a probability distribute over two categories (is adopted by current proposal vs. not).
The architecture is trained end-to-end with a multi-task loss.}}
  \label{fig:base_fast_rcnn}
\end{figure*}
\fi

\subsection{Parameter analyses and ablation studies}
\label{subsec:param}
%% 在推理中， Unary term中  每个零件 的语义概率分布中，考虑部件假设的数目

\paragraph{\textbf{Parameter $K^c$}}
When performing component inference and labeling (Section~\ref{sec:labelling}),
the number of top-ranked part hypotheses, denoted by $K^c$, selected for each component $c$ in defining
the unary potential (Equation (\ref{eq:CRF_unary_P}))
is an important parameter of our method.
We experiment the parameter settings $K^c$ being set to $1, 3, 5, 10$ and $all$ respectively,
while keeping all other parameters unchanged.
$all$ means to use all part hypotheses of $c$ (i.e., $K^c = H^c$).
The results of per-category average IoU are shown in row 16-20 of Table~\ref{tab:LabellingAssessment}.
For object categories, the best performance is obtained when using $K^c=all$ for each component.
For scene categories (the last two columns), however, $K^c<all$ leads to better performance.
This is because, for scene categories, the top ranked hypotheses, corresponding to the early groupings
emerged in the hierarchical sampling process, are usually the individual objects in the scene.
Such groups occur more frequently and hence more reliable to capture.
The subsequent groupings, however, imply larger scale, inter-object structures.
Since the spatial relationships between objects are usually loose, as we have pointed out earlier,
such structures are less reliable (hard to learn), especially when the grouping scale becomes very large.

\if 0
\begin{figure}[!ht]
  \centering
  \includegraphics[width=\linewidth]{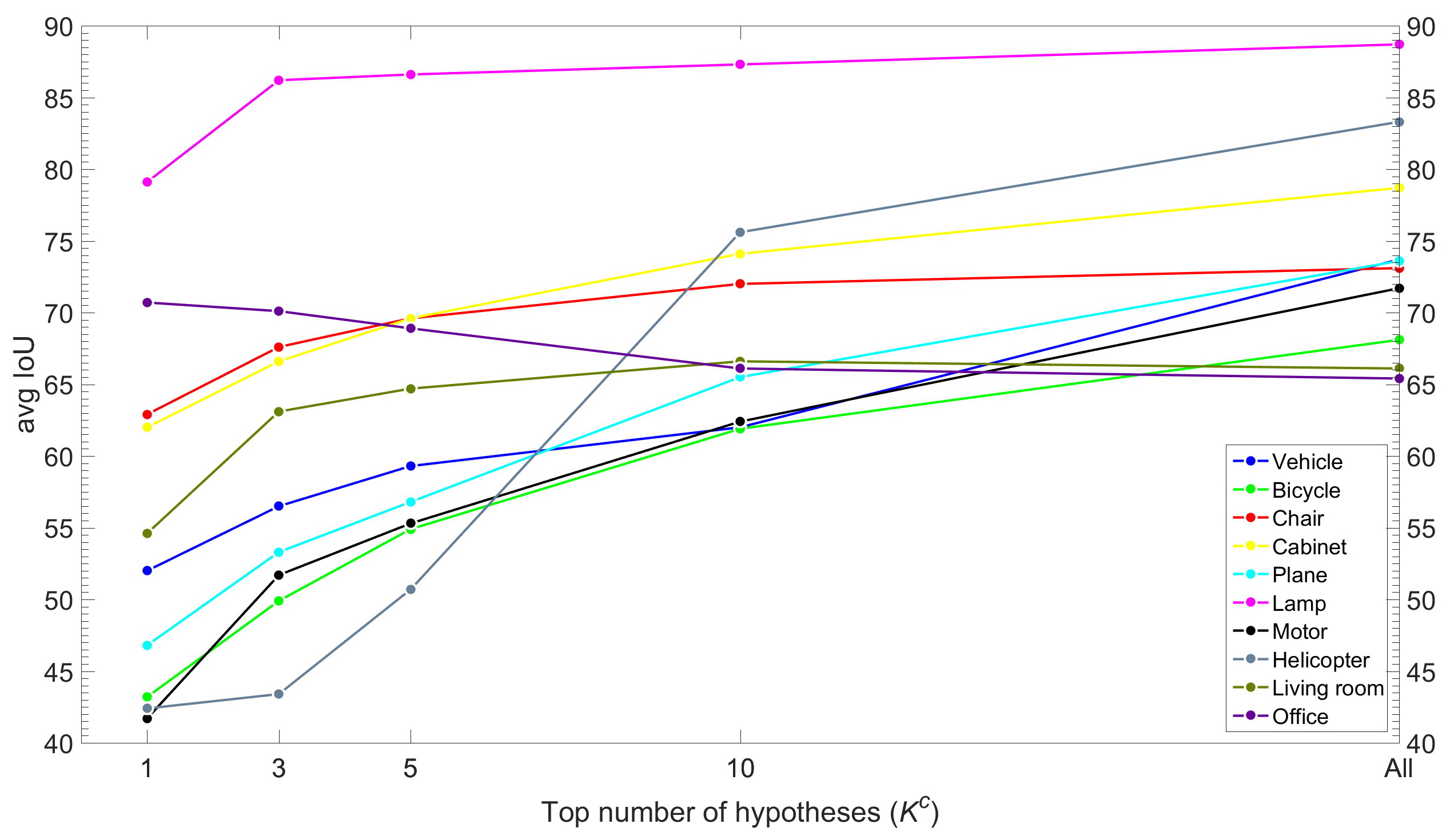}
  \caption{Labeling accuracy (average IoU) over different choices of $K^c$.}
  \label{fig:PerfVsPara}
\end{figure}
\fi

\begin{figure}[!t]
  \centering
  \includegraphics[width=\linewidth]{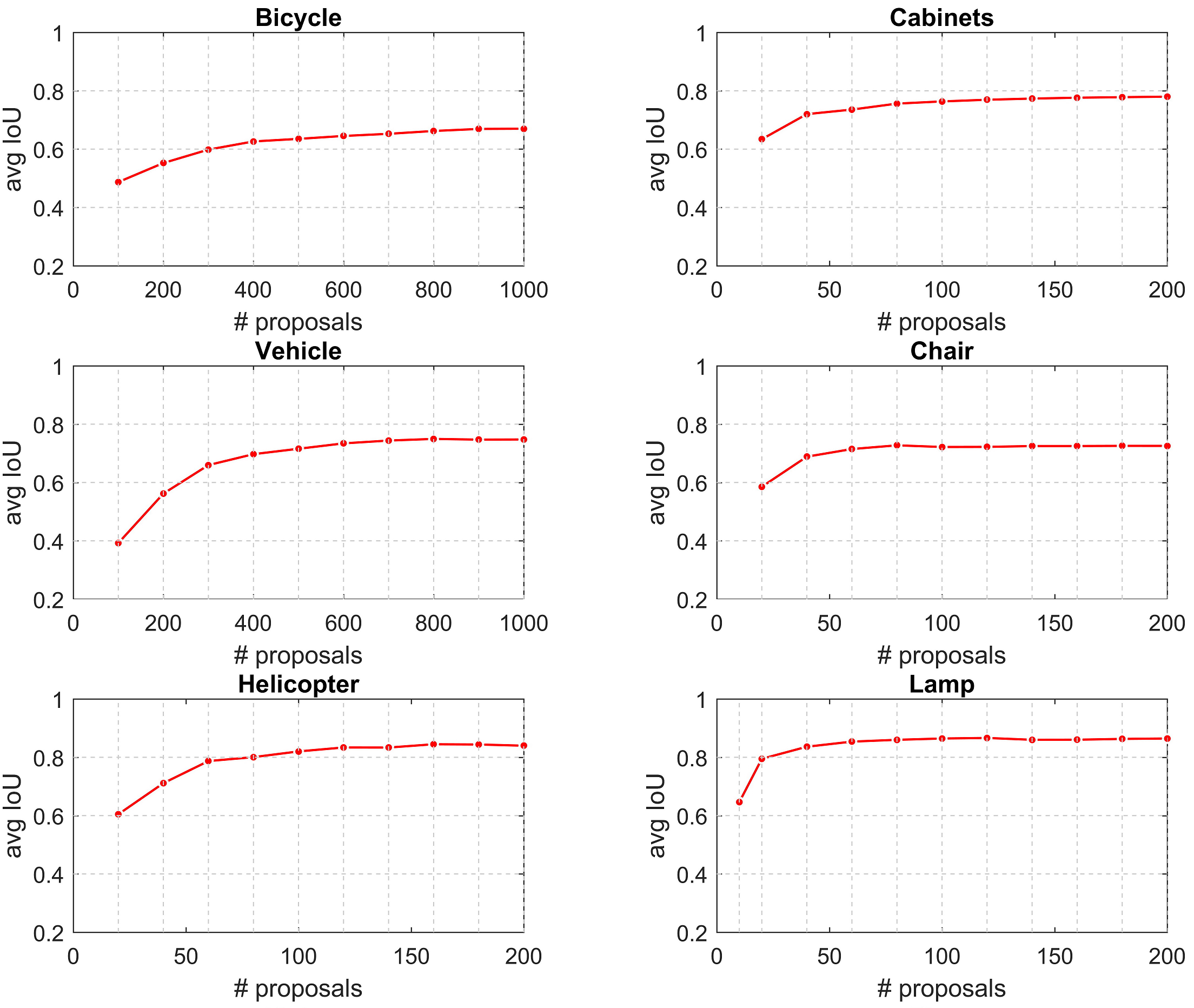}
  \caption{\new{Labeling accuracy (average IoU) vs. number of part hypotheses.}}
  \label{fig:Performance_vs_num}
\end{figure}

%% proposal数目，对最终的性能的影响。 这儿还缺一个实验待补充
\new{
\paragraph{\textbf{Labeling performance over part hypothesis count}}
We also evaluate our method with the varying number of part hypotheses generated.
Figure~\ref{fig:Performance_vs_num} shows the plots of avg IoU over the number of
part hypotheses.
\kx{The test is performed on six object categories of our benchmark dataset and
the results on more categories can be found in the supplementary material.
The same goes for all the plots shown in this paper.}
Generally, the performance grows as the number of hypotheses increases,
but stops growing at a specific number.
For all categories, we choose the hypothesis count no greater than $1000$,
even for structurally complicated categories such as vehicle and bicycle.
This shows that
our approach is insensitive to the initial number of part hypotheses.}

%% 在推理中， 对于每个部件假设，仅考虑分类
\new{
\paragraph{\textbf{Labeling performance without confidence score}}
For each part hypothesis, a confidence score is regressed by our network, which measures
how likely it represents an independent semantic part.
This score is employed in defining the unary potential in the global labeling inference.
To test its effect,
we experiment an ablated version of our method without considering this confidence score
(by setting $s_i^c=1$ in Equation (\ref{eq:CRF_unary_P})), while keeping all other parameters unchanged.
The experimental results are reported in row 13 of Table~\ref{tab:LabellingAssessment}.
For all categories, our method works better when incorporating confidence score.
In particular, the improvement of average IoU over `w/o score' ranges from $0.7\%$ to $5.3\%$.}

\subsection{Evaluation on part hypothesis generation}
\label{subsec:hypo}
\begin{figure}[!t]
  \centering
  \includegraphics[width=\linewidth]{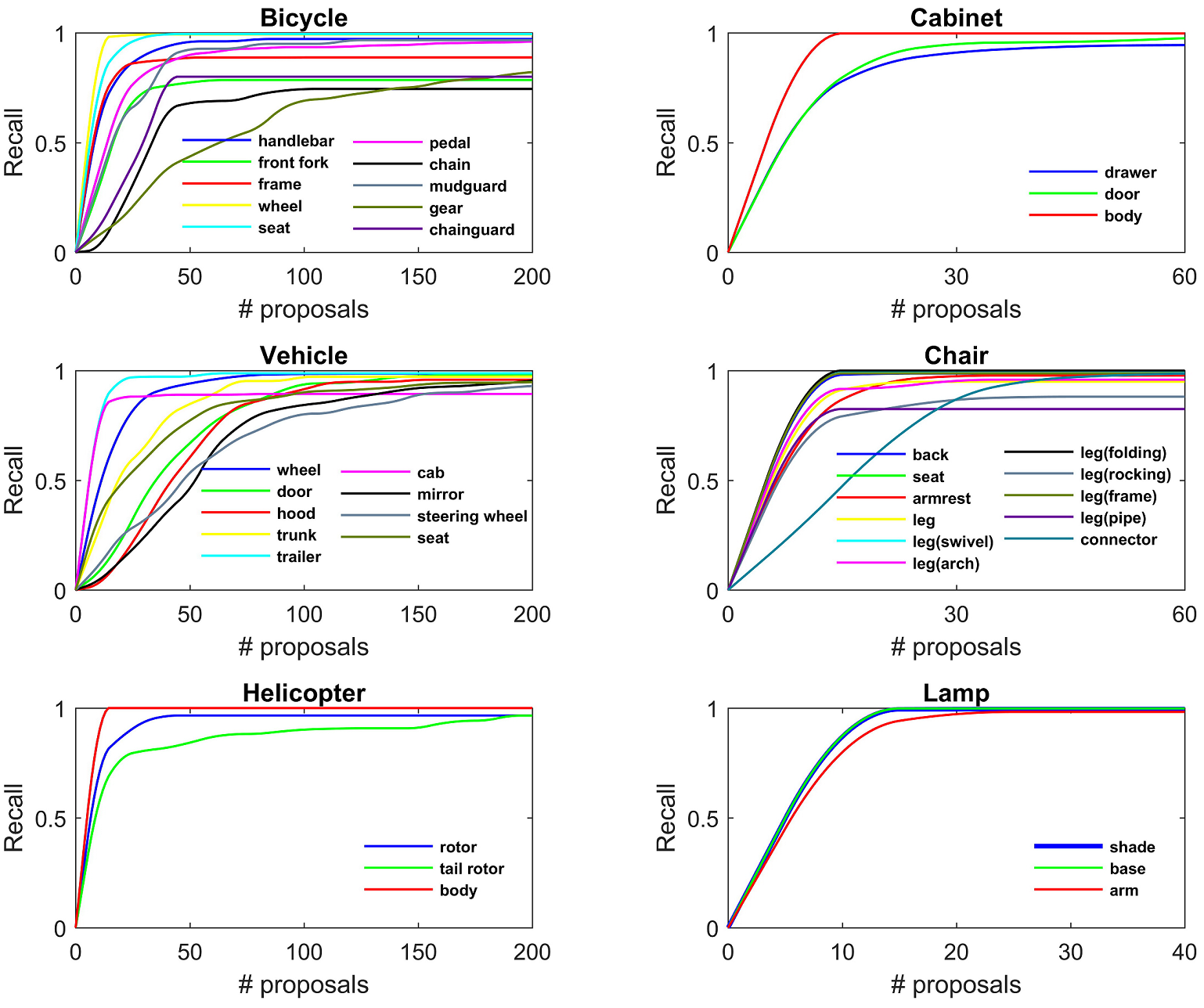}
  \caption{Recall rate on semantic parts over varying number of part hypotheses, when IoU against ground-truth is fixed to $50\%$, over six categories.}
  \label{fig:Perf_recall_vs_num}
\end{figure}

\new{
\paragraph{\textbf{Part hypothesis quality vs. hypothesis count}}
%despite being fully heuristic(no machine learning involved).
%Overall, the algorithm gradually rise recall as the number of part hypotheses increases.
%When reaching a certain number, the algorithm will provide stable results.
In Figure~\ref{fig:Perf_recall_vs_num}, we study part hypothesis quality over the change of
the number of sampled hypotheses.
Part hypothesis quality is measured as follows:
For the sampled hypotheses whose IoU is greater than $50\%$,
we compute their recall rate over the ground-truth semantic parts.
We find that the hypothesis quality (recall rate) grows rapidly as the number of hypotheses increases,
becomes stable fast at a moderate hypothesis count.
For complex categories (e.g., bicycle, vehicle, motor, office), the count is lower smaller $600$,
For other categories (e.g., cabinets, chair, lamp, living room), on the other hand,
the number is no greater $200$.
These numbers show that our sampling algorithm produces high quality hypotheses
with a moderate sampling size, much smaller than that of exhaustive enumeration.}   %end

\begin{figure}[!t]
  \centering
  \includegraphics[width=3.4in]{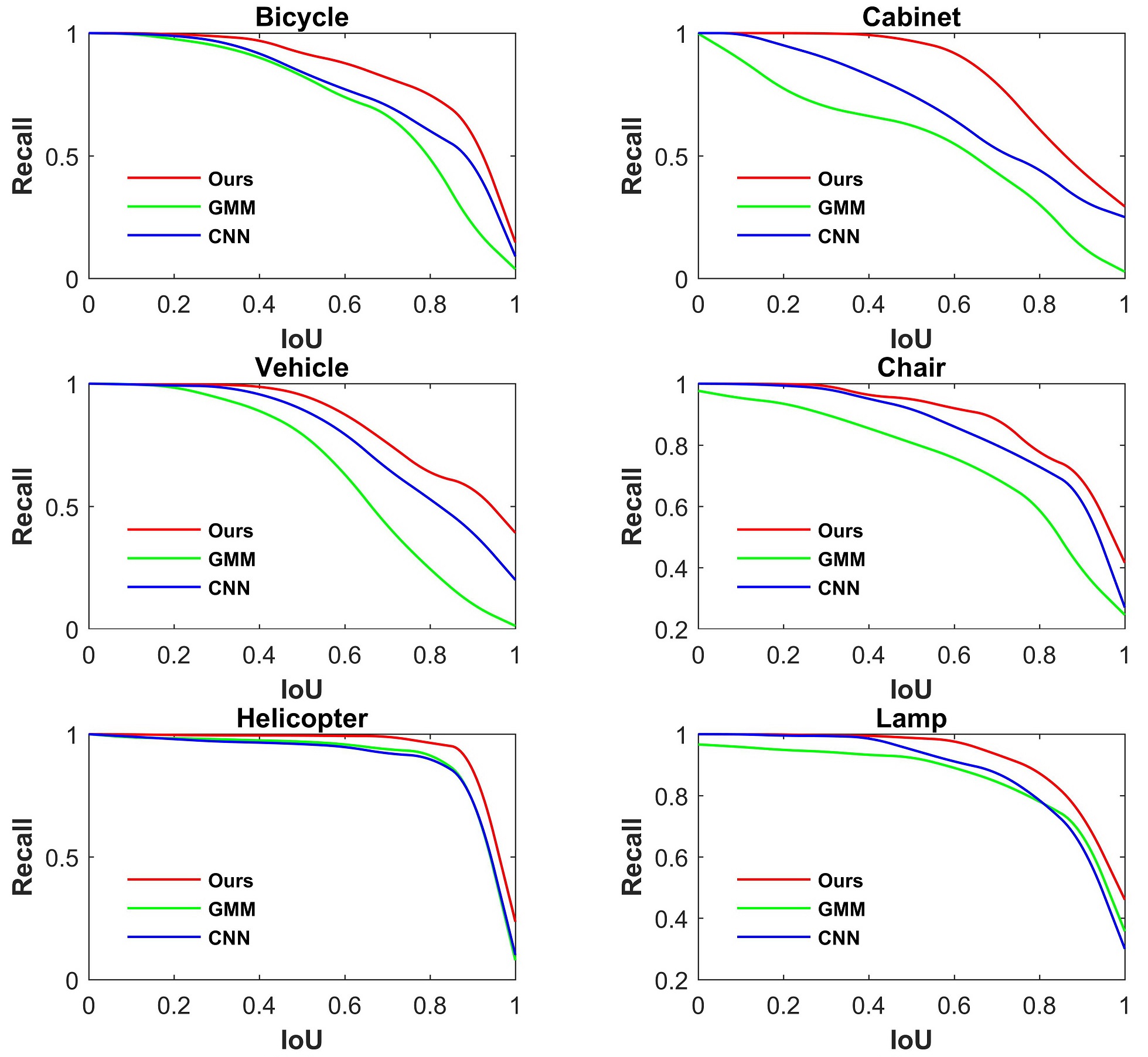}
  \caption{\wxg{Performance (recall rate vs. average IoU) comparisons between our hierarchical grouping algorithm and the two alternative methods (GMM- and CNN-based hypothesis generation) over six categories.}}
  \label{fig:Comp_GMM_2}
\end{figure}

\paragraph{\textbf{Comparison to alternatives}}
We assess the quality of part hypotheses
by comparing our hierarchical grouping algorithm with two alternative methods.
The first method is the CNN-based hypothesis generation we have mentioned above.
The second one learns for each shape category
a Gaussian Mixture Model (GMM) modeling the position and scale distribution
of bounding boxes of semantic parts.
Given an input shape, the method generates part hypotheses
by sampling the GMM of the corresponding shape category.
%An example of GMM based hypothesis generation is shown in the inserted figure.
%\begin{wrapfigure}{r}{0.4\linewidth}
%\vspace{-0.45cm}
%\hspace{-0.7cm} \includegraphics[width=1.18\linewidth]{Images/Comp_GMM_1}
%\vspace{-0.7cm}
%\end{wrapfigure}
To make the comparison, we generate the same number of top hypotheses for all methods.
While our method samples the top number of hypotheses according to the hierarchical sampling order
(Section~\ref{sec:sampling}), GMM samples based on probability.
For CNN-based method, we use all hypotheses generated.
We plot in Figure~\ref{fig:Comp_GMM_2} the curves of recall rate over average IoU for the three methods.
It can be observed that our method produces the highest quality part hypotheses.
The GMM-based method is a probabilistic sampling based approach which is fuzzy and cannot produce hypotheses with accurate boundaries.

%\wxg{\paragraph{\textbf{Comparison to baseline (CNN-based hypothesis generation)}}
%In Figure~\ref{fig:Comp_GMM_2}, we also compare our hierarchical grouping algorithm with CNN-based hypothesis generation baseline method.
%In essence, it is a method to extract the part hypothesis by using the multi-scale boxes, and it cannot handle the boundary problem well.
%At the same time, small components occupy less volume in voxel space,
%therefore, the ability of feature expression is limited for small components.
%It is easy to misclassify small components in refinement branch.
%The network architecture and its detailed explanation can be found
%in the supplemental material.}

%given in Figure~\ref{fig:Comp_GMM_1}.
%\begin{figure}[!b]
%  \centering
%  \includegraphics[width=0.6\linewidth]{Images/Comp_GMM_1}
%  \caption{Part hypotheses (visualized by bounding boxes) of bicycle chain, generated by the GMM-based method.}
%  \label{fig:Comp_GMM_1}
%\end{figure}
%

\subsection{Network analysis}
To evaluate our network design, we study the effect of each of the three towers,
\emph{local}, \emph{global}, and \emph{contextual}, of our multi-scale CNNs.
Specifically, we train and test networks with two different combinations, `\emph{local} only' and `\emph{local}+\emph{global}'.
In Table~\ref{tab:LabellingAssessment} (row 14 and 15), we show the results when different combinations of the towers are employed.
Our multi-scale CNNs architecture (with all three towers) results in the best performance
for object categories,
while the `\emph{local}+\emph{global}' architecture leads to higher performance for scene categories.
This can be explained, again, by the fact that indoor scenes, even from the same category, have loosely defined structure
and possess much layout variation. This makes it difficult to learn its global structure reliably, even
with the help of \emph{contextual} tower, when training data is not sufficiently large.
Since our method is not targeted to scene parsing, we leave this for future work.
Nevertheless, our method still obtains acceptable accuracy for scenes, verifying the ability of our method in
handling structures across a large range of scales.
%\todo{This is because of that
%our multi-scale CNNs architecture has similarity for object categories,
%and contextual information has a certain regularity.
%But for scene categories, the law of overall layout maybe not exist, or we can not find it from relative small training data.}

\if 0
\begin{table}[!ht]
\caption{Labeling accuracy (avg IoU) comparisons among our multi-scale CNNs architecture (\textsl{`Ours'} for short), \textsl{`Local Only'} architecture and   \textsl{`Local+Global'} architecture.}
\centering
  \begin{tabular}{lccc}
    \hline
      & Vehicle &  Carbinet & Office \\
    \hline
    \textsl{Local Only}   & 50.4 & 68.6 & 54.8 \\
    \textsl{Local+Global} & 69.2 & 75.4 & \textbf{76.4} \\
    \textsl{Ours}         & \textbf{73.7} & \textbf{78.7} & 65.4 \\
    \hline
  \end{tabular}
\label{tab:NetworkAnalysis}
\end{table}
\fi

\subsection{Training and testing time}
\new{
Our implementation is built on top of the Caffe~\citep{Jia_2014} framework based on the standard settings.
We used Adam~\citep{Kingma_2014} stochastic optimization for training with a mini-batch of size $64$.
The initial learning rate is $0.001$.
The numbers of training samples are listed in Table~\ref{tab:LabellingAssessment} (Row 5).
Training takes about $40$ minutes per $1K$ iteration, and about $13.3$ hours per shape category.
Testing our CNN network consumes about $0.02$ seconds per part hypothesis.
The whole task takes about $20$ seconds per shape.
Table~\ref{tab:timing} shows the timing for various algorithmic components.
Runtime computations were performed using a Nvidia GTX 1080 GPU and a 4 core Intel i7-5820K CPU machine.}

\begin{table}[!ht]
\caption{\new{Timing (in second) of various algorithmic components.}}
\centering
  \begin{tabular}{lrr}
    \hline
     Technical component & Objects &  Scenes \\
    \hline
    \textsl{Hypothesis generation}   & 5.2 & 5.6 \\
    \textsl{CNN testing} & 10.5 & 4.5 \\
    \textsl{Higher-order CRF}         & 7.4 & 3.6 \\
    \hline
  \end{tabular}
\label{tab:timing}
\end{table}

\subsection{An application to shape correspondence}
\new{
We apply our approach to component-level shape correspondence, which has been studied in~\cite{Alhashim_2015}
and~\cite{Zhu_2017}.
Given two multi-component shapes, we first use our method to group and label the components for each shape.
Based on the semantic labeling, we find a global alignment for the local canonical frames of the two shapes.
This is achieved by minimizing the spatial distance between every two components with the same label,
each from one of the two shapes.
Given a pair of semantic parts with the same label, each from one of the two shapes being matched,
we align their bounding boxes and find correspondence for their enclosed components.
In particular, we find for each component from one shape
the spatially closest counterpart from the other shape.
After the bidirectional search and a post-processing of conflict resolving (always keep the closer
one if there are multiple matches), we return for each component from one shape
a unique matching component from the other shape.

We compare our simple method with the two state-of-the-arts, \cite{Alhashim_2015} and~\cite{Zhu_2017},
on the benchmark dataset GeoTopo~\cite{Alhashim_2015}.
Table~\ref{tab:ShapeCor} reports the results of precision and recall for component matching.
Our method achieves comparable results to theirs, and performs better on categories with
higher number of semantic parts, such as airplane and velocipedes.
Note that our method does not consider any high-level structural information such as symmetry.}

\begin{table}[!ht]
\caption{
\new{Comparisons on precision and recall rates against two state-of-the-art shape correspondence methods~\citep{Alhashim_2015} and ~\citep{Zhu_2017}.
}}
\centering
  \begin{tabular}{lccccccc}
    \hline
    \multirow{2}{*}{Category} & \multicolumn{2}{c}{GeoTopo} &  \multicolumn{2}{c}{DDS} & \multicolumn{2}{c}{Our} \\\cline{2-7}
    & {P} & {R} & {P} & {R} & {P} & {R} \\
    \hline
    \textsl{Chair}  & 0.69  & 0.67 & \textbf{0.83} & \textbf{0.83} & 0.71 & 0.75   \\
    \textsl{Table}  & 0.63 & 0.61 & \textbf{0.81} & \textbf{0.86} & 0.79 & 0.80\\
    \textsl{Bed}   & 0.60 & 0.62 & \textbf{0.78} & \textbf{0.81} & 0.75 &0.72\\
    \textsl{Airplane} & 0.60 &0.68 & 0.80 & 0.85 & \textbf{0.85} & \textbf{0.91} \\
    \textsl{Velocipedes}    & 0.47 & 0.44 & 0.43 & 0.49 & \textbf{0.48} & \textbf{0.55} \\
    \hline
  \end{tabular}
\label{tab:ShapeCor}
\end{table}

\section{Conclusion}
\new{We have studied a new problem of labeled segmentation of off-the-shelf
3D models based on the pre-existing, highly fine-grained
components. We approach the problem with a novel solution
of part hypothesis analysis.
The core idea of our approach is exploiting part hypothesis as a mid-level representation
for semantic composition analysis of 3D shapes.
This leads a highly robust labeling algorithm which can handle highly complicated structures
in various scales and forms.
Our work contributes, to the best of our knowledge, the first component-wise labeling algorithm
that simultaneously works for single objects and compound scenes.

\new{
The success of our method is due to \emph{three key features}:
\emph{First}, part hypotheses are generated in a principled way, based on a bottom-up
hierarchical grouping process, guided by three intuitive criteria.
\emph{Second}, a deep neural network is trained to encode part hypothesis, rather than components,
accounting for both local geometric and global contextual information.
\emph{Third}, the higher order potential in our CRF-based formulation adopts a soft consistency
constraint, providing more degree of freedom in optimal labeling search.
}

\begin{figure}[t]
  \centering
  \includegraphics[width=\linewidth]{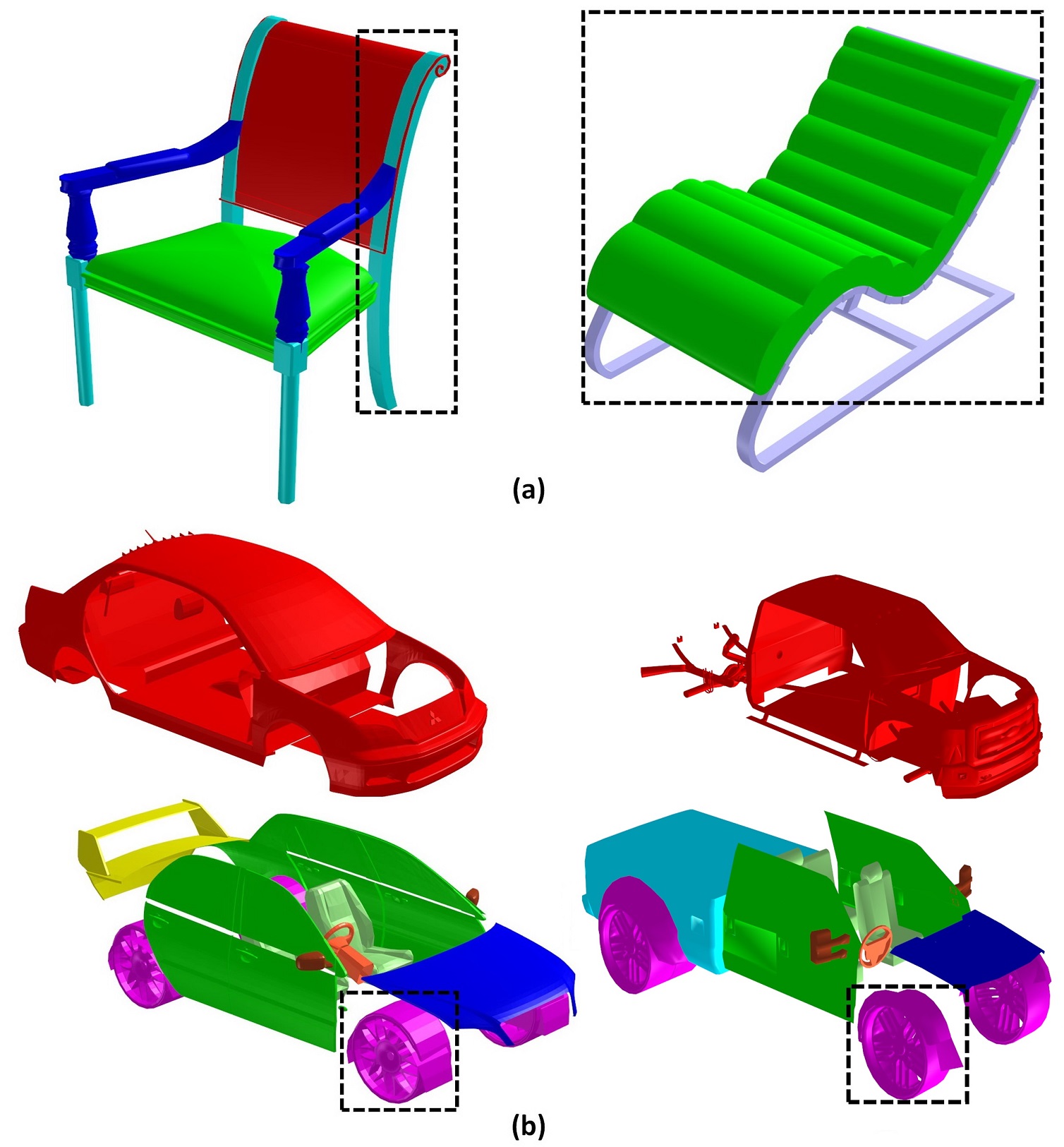}
  \caption{\wang{There are two typical kinds of failure cases.
  (a):
  Our method cannot handle the case where the components are under-segmented with respect to semantic parts.
  In the left chair, the component marked in the dashed box encompasses both leg and back parts. The seat and back part of the right chair are merged as a single component. For such examples, a correct labeling cannot be obtained without a further breaking of this component.
  (b):
  When the bounding boxes of two part hypotheses overlap significantly, due to, for example, shape concavity, the labeling of them can be misled by each other. For example, the concave mudguard can be mistakenly labeled as wheel, due to its bounding box overlap against the wheel.}}
  \label{fig:failure}
\end{figure}

\paragraph{\textbf{Limitations, failure cases and future works}}
Our approach has several limitations, which point out directions for future investigation.
\emph{First}, our current solution only groups the components but not further segment them,
it thus cannot handle the case where the components are under-segmented with respect to semantic parts.
Figure~\ref{fig:failure}(a) shows two examples of such failure case.
For such examples, a correct labeling cannot be obtained without a further breaking of this component.
According to our statistics, only about $6\%$ shapes in ShapeNet have such issue, \wang{based on our own set of semantic labels}.
As a future work, we would consider incorporating component-level segmentation
into our framework.
%Meanwhile, our method requires some pre-processing and post-processing, which also limits its application.
%
\wang{Figure~\ref{fig:failure}(b) shows another type of failure case. When the bounding boxes of two part hypotheses overlap significantly, due to, for example, shape concavity, their labeling can be misleading.}
\wang{Currently, our method does not produce hierarchical part grouping and labeling, as in~\cite{Yi_SG17}.
It would be interesting to investigate extending our hypothesis analysis for the task of hierarchical segmentation.
For example, how to determine the order of grouping, or the structure of the hierarchy, is a non-trivial task.}
\wang{Another worthy topic for future research is the integration of CRF in the
deep neural networks to make the entire model end-to-end trainable while avoiding relying on
strong supervision~\cite{Kalogerakis_CVPR17}.}

\begin{figure*}
  \centering
  \includegraphics[width=\textwidth]{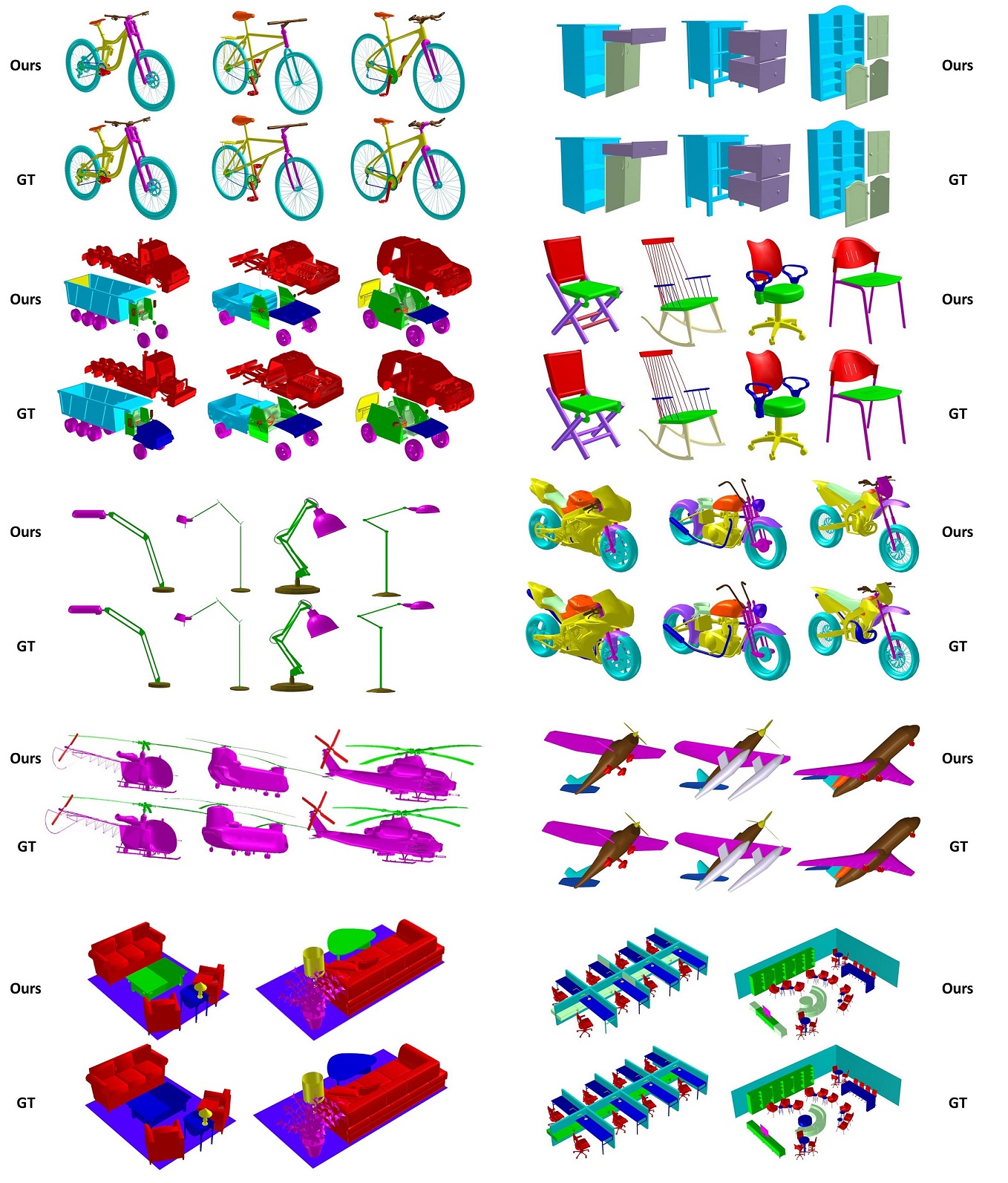}
  \caption{A gallery of semantic labeling results on raw 3D models with complicated structure. `GT' denotes ground-truth labeling annotated manually.}
  \label{fig:MoreResults}
\end{figure*}
}

\section*{Acknowledgements}
We thank the anonymous reviewers for their valuable comments.
The authors are grateful to Hao Su for fruitful discussion, and Zheyuan Cai and Yahao Shi for the help on data preparation.
This work was supported in part by NSFC (61572507, 61532003, 61622212, 61502023 and U1736217).
Kai Xu is also supported by a visiting research scholarship offered by China Scholarship Council (CSC) and Princeton University.

\bibliographystyle{ACM-Reference-Format}
\bibliography{cadlabel}

\end{document}